\documentclass{aa}  
\usepackage{graphicx}
\usepackage{txfonts}

\newcommand{\mm}{\,{\rm mm}}
\newcommand{\mum}{\,{\mu {\rm m}}}

\newcommand{\as}{^{\prime\prime}}

\newcommand{\bdm}{\begin{displaymath}}
\newcommand{\edm}{\end{displaymath}}
\newcommand{\beq}{\begin{equation}}
\newcommand{\eeq}{\end{equation}}
\newcommand{\bit}{\begin{itemize}}
\newcommand{\eit}{\end{itemize}}
\newcommand{\ben}{\begin{enumerate}}
\newcommand{\een}{\end{enumerate}}
\newcommand{\bfi}{\begin{figure}[htb]}
\newcommand{\bpfi}{\begin{figure}[p]}

\newcommand{\ha}{$\rm H\alpha$}
\newcommand{\hb}{$\rm H\beta$}

\begin{document}

\title{The Near-Infrared Spectrograph (NIRSpec) on the James Webb Space Telescope}
\subtitle{III. Integral-field spectroscopy}
   \titlerunning{NIRSpec integral-field spectroscopy}


   \author{T. B\"oker \inst{1}
    \and S. Arribas\inst{2}
    \and N. L\"utzgendorf\inst{1}
    \and C. Alves de Oliveira\inst{3}
    \and T. L. Beck\inst{4}
    \and S. Birkmann\inst{1}
    \and A. J. Bunker\inst{5}
    \and S. Charlot\inst{6}
    \and G. de Marchi\inst{7}
    \and P. Ferruit\inst{3}
    \and G. Giardino\inst{8}
    \and P. Jakobsen\inst{9}
    \and N. Kumari\inst{10}
    \and M. L\'{o}pez-Caniego\inst{11}
    \and R. Maiolino\inst{12}
    \and E. Manjavacas\inst{10}
    \and A. Marston\inst{3}
    \and S. H. Moseley\inst{13}
    \and J. Muzerolle\inst{4}
    \and P. Ogle\inst{4}
    \and N. Pirzkal\inst{10}
    \and B. Rauscher\inst{14}
    \and T. Rawle\inst{1}
    \and H.-W. Rix\inst{15}
    \and E. Sabbi\inst{4}
    \and B. Sargent\inst{4}
    \and M. Sirianni\inst{1}
    \and M. te Plate\inst{1}
    \and J. Valenti\inst{4}
    \and C. J. Willott\inst{16}
    \and P. Zeidler\inst{10}
          }
\authorrunning{B\"oker et al.}

\institute{ 
European Space Agency, Space Telescope Science Institute, Baltimore, Maryland, USA \and
Centro de Astrobiolog\'ia, (CAB, CSIC--INTA), Departamento de Astrof\'\i sica, Madrid, Spain \and
European Space Agency, European Space Astronomy Centre, Madrid, Spain \and
Space Telescope Science Institute, Baltimore, Maryland, USA \and
Department of Physics, University of Oxford, United Kingdom \and
Sorbonne Universit\'{e}, CNRS, UMR 7095, Institut d’Astrophysique, Paris, France \and
European Space Agency, European Space Research and Technology Centre, Noordwijk, The Netherlands \and
ATG Europe for the European Space Agency, European Space Research and Technology Centre, Noordwijk, The Netherlands \and
Cosmic Dawn Center, Niels Bohr Institute, University of Copenhagen, Denmark \and
AURA for the European Space Agency, Space Telescope Science Institute, Baltimore, Maryland, USA \and
Aurora Technology for the European Space Agency, European Space Astronomy Centre, Madrid, Spain \and
Kavli Institute for Cosmology, University of Cambridge, United Kingdom \and
Quantum Circuits, Inc., New Haven, Connecticut, USA \and
NASA Goddard Space Flight Center, Greenbelt, Maryland, USA \and
Max-Planck Institute for Astronomy, Heidelberg, Germany \and
NRC Herzberg, Victoria, British Columbia, Canada
}

   \date{Received tbd; accepted tbd}

 
  \abstract{
The Near-Infrared Spectrograph (NIRSpec) on the James Webb Space Telescope (JWST) offers the first opportunity to use integral-field spectroscopy from space at near-infrared wavelengths. More specifically, NIRSpec's integral-field unit can obtain spectra covering the wavelength range $0.6 - 5.3\mum$ for a contiguous $3.1\as \times 3.2\as$ sky area at spectral resolutions of $R \approx 100$, 1000, and 2700. In this paper we describe the optical and mechanical design of the NIRSpec integral-field spectroscopy mode, together with its expected performance. We also discuss a few recommended observing strategies, some of which are driven by the fact that NIRSpec is a multipurpose instrument with a number of different observing modes, which are discussed in companion papers. We briefly discuss the data processing steps required to produce wavelength- and flux-calibrated data cubes that contain the spatial and spectral information. Lastly, we mention a few scientific topics that are bound to benefit from this highly innovative capability offered by JWST/NIRSpec.}


   \maketitle
%
\section{Introduction}

Due to its large collection area, cryogenic temperature, and location above the Earth's atmosphere, the James Webb Space Telescope (JWST) will expand the boundaries of near-infrared (NIR) and mid-infrared (MIR) astronomy\footnote{Throughout this paper, the wavelength ranges are defined as $1 - 5\mum$ (NIR) and $5-30\mum$ (MIR).}. To take full advantage of its unique capabilities, JWST relies on an innovative suite of science instruments designed to advance a rich list of scientific topics \citep[see][]{Gar06}. The Near-Infrared Spectrograph (NIRSpec), in particular, offers two observing modes that have never before been used for NIR and MIR astronomy from space: a microshutter slit-selection mechanism for multi-object spectroscopy and an integral-field spectroscopy (IFS) mode that employs image-slicing optics contained in a so-called integral-field unit (IFU).
While the IFS mode is the topic of this paper, three accompanying articles provide further details of NIRSpec's scientific goals and overall design \citep[][hereafter Paper I]{Jak21}, its multi-object spectroscopy capabilities \citep[][Paper II]{Fer21}, and the prospects for carrying out fixed slit spectroscopy of exoplanets \citep[][Paper IV]{Bir21}.

Integral-field spectroscopy is a relatively recent addition to the toolbox of observational astronomy. The advantages of obtaining full spectral information over a contiguous two-dimensional field of view in a single exposure are obvious, but the technical challenges of optically rearranging a two-dimensional field of view into a one-dimensional ``long slit'' without losing spatial resolution were only overcome about two decades ago. At visible wavelengths, IFS instruments have used a variety of techniques such as image slicing \citep[e.g., MUSE;][]{Bac10}, arrays of microlenses \citep[e.g., TIGER or SAURON;][]{Bac95,Bac01}, optical fibers \citep[e.g., SILFID/ARGUS or INTEGRAL;][]{Van88, Arr98}, or a combination of microlenses and fibers in instruments such as GMOS \citep{All02} or VIMOS \citep{LeF03}. 

\begin{table}[htb]
\caption{High-level characteristics of the NIRSpec IFS mode.}
\label{tab:ifs_summary}      
\centering                          
\begin{tabular}{l l}        
\hline                
   field of view & $3.1\as \times 3.2\as$ \\
   no. of slices & 30 \\      
   spaxel size$^1$ & $0.103\as \times 0.105\as$ \\
   wavelength range$^2$ & 0.6 - $5.3\mum$ \\
   spectral resolution  & $\approx$100, 1000, or 2700 \\
\hline                                   
\end{tabular}
\tablefoot{\scriptsize $^1$ Defined as (slice width $\times$ pixel size). $^2$ For the NIRSpec gratings, coverage of the full wavelength range requires multiple exposures with different combinations of grating and filter (see Table\,\ref{tab:gap} for details).}
\end{table}

At infrared (IR) wavelengths, however, the use of optical fibers is challenging, especially for wavelengths beyond $\lambda = 1.9\mum$, mostly because fiber materials that efficiently transmit IR wavelengths are not readily available. Therefore, most IR IFUs deployed on ground-based telescopes use image slicing optics, for example
3D \citep{Wei96}, SINFONI \citep{Eis03}, NIFS \citep{McG03}, and KMOS \citep{Sha06}, although some use a microlens array \citep[e.g., OSIRIS;][]{Lar06}. The combination of some of these instruments with adaptive optics (AO) systems has significantly improved the angular resolution imposed by the atmospheric seeing. Today, IFS is a mature technique for ground-based astronomy, and all future extremely large telescopes (ELTs) have plans to include IFUs in their first generation of instruments, for example ELT/HARMONI, \citep{Tha20}, GMT/GMTIFS \citep{McG12}, and TMT/IRIS \citep{Moo14}. Despite the overall progress, ground-based, AO-assisted, IFS observations will remain challenging due to the limited availability of natural guide stars and the technical challenges associated with using lasers to create artificial guide stars. Even the best AO systems suffer from persistent variations in the quality of the seeing correction, which can result in a nonoptimal Strehl ratio and an unstable point-spread function (PSF), both of which often vary with time and/or position in the field of view. 

   \begin{figure*}[t]
   \centering
   \includegraphics[width=0.85\hsize]{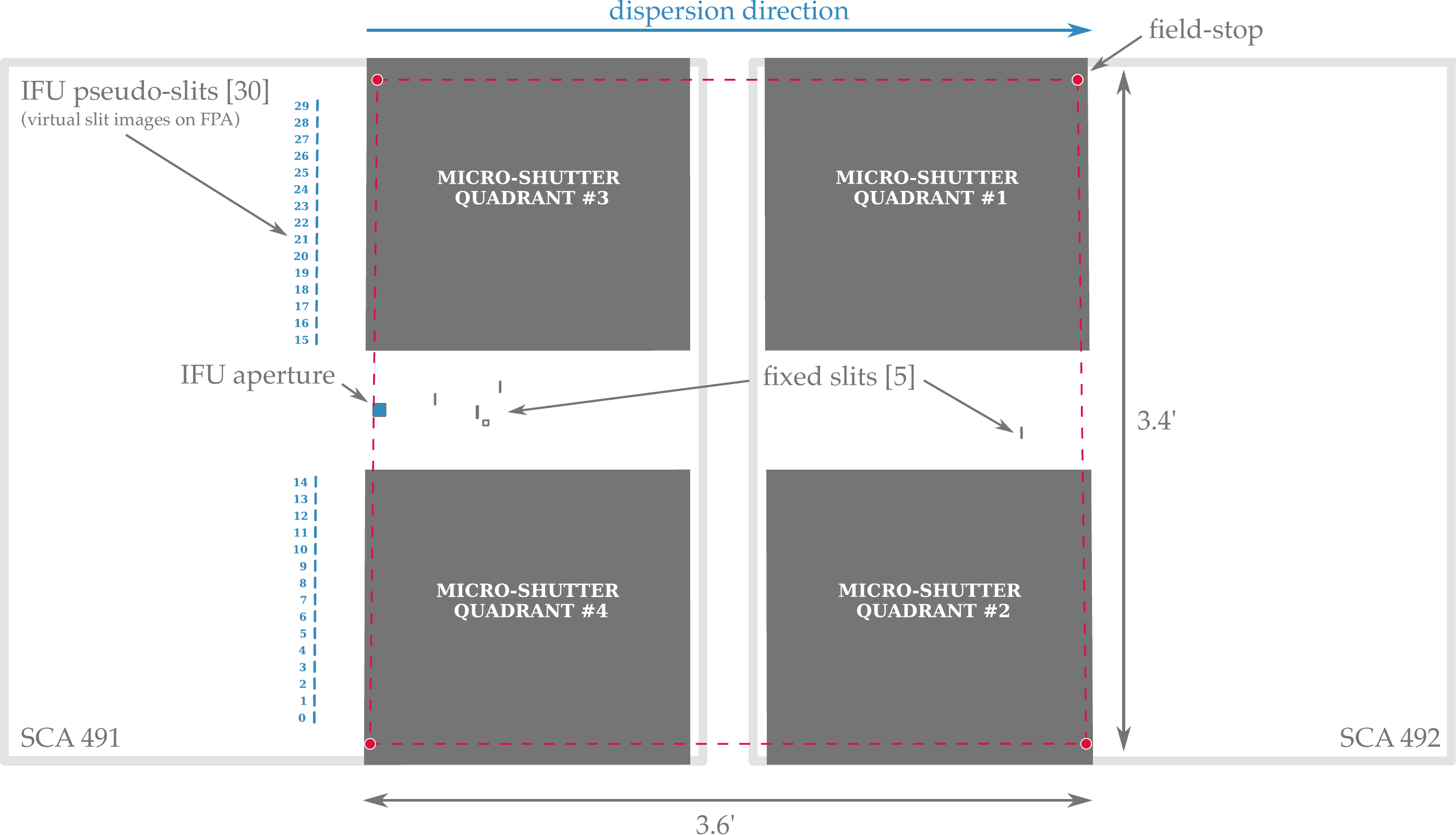}
      \caption{Geometry of the NIRSpec field of view. In the entrance focal plane, it is limited by the field stop (dashed red line), which is slightly undersized with respect to the microshutter array (dark gray filled squares). The metal frame holding the four quadrants has cutouts for the fixed slits and the IFU aperture. The two sensor chip assemblies (SCAs) populating the NIRSpec focal plane assembly (FPA) are over-plotted (large light gray open squares). Also shown are the locations of the 30 undispersed slice images on the detector, i.e., where they would appear if the MIRROR position were used instead of a disperser. It should be noted that 
the IFU spectra fall onto the same detector area as those of open microshutters, which is why the IFS and MOS modes of NIRSpec are, in general, mutually exclusive (see discussion in Sect. \ref{subsec:position}).
              }
         \label{fig:fov}
   \end{figure*}

Therefore, use of the IFU technology in space-based instruments has obvious advantages given the continuous wavelength coverage and inherent PSF stability offered by avoiding the Earth's atmosphere. However, progress has been slow, mostly due to the long timescale and limited flexibility associated with space missions, which makes it difficult to incorporate relatively recent techniques. The first space-based IFU, designed for far-IR wavelengths, was successfully used on the PACS instrument on board the Herschel Observatory \citep{Pog10}. The JWST will be the first mission to employ this technology at NIR and MIR wavelengths, with two\footnote{The IFS mode of the MIR instrument (MIRI) is described in \cite{wel15}.} science instruments offering an IFS mode. This represents a major step forward because, in addition to the inherent advantages of IFS, these observations will benefit from the unprecedented sensitivity, the wide and continuous spectral range, and the stable PSF provided by the JWST.

The topic of this paper is the IFS mode of NIRSpec, whose high-level characteristics are summarized in Table\,\ref{tab:ifs_summary}. Section \ref{sec:design} describes the design and main characteristics of the NIRSpec IFU. The scientific performance of the IFS mode, which results from the combination of the IFU itself with other elements of NIRSpec and the observatory, is discussed in Sect. \ref{sec:perf}, while Sect. \ref{sec:strategies} describes different observing strategies tailored for the optimal subtraction of background and leakage through the microshutter assembly (MSA). Section\,\ref{sec:pipeline} briefly describes the data reduction pipeline used to produce calibrated IFS data cubes. A few examples of scientific use cases are presented in Sect. \ref{sec:science}, and a brief summary is provided in Sect. \ref{sec:summary}.

\section{The NIRSpec IFU assembly}\label{sec:design}

\subsection{Position within the optical path}\label{subsec:position}
NIRSpec is a multipurpose instrument, and in order to allow for the various observing modes within a reasonably compact instrument design, they mostly share the same optical path and detector real estate. This is illustrated in Fig.\,\ref{fig:fov}, which shows the layout of the NIRSpec field of view, superposed on the area covered by the two NIRSpec detectors. The four quadrants of the MSA \citep{Fer21} are mounted on a metal frame that has cutouts for the five (permanently open) fixed slits and the IFU aperture. The locations of the 30 undispersed slice images on the detector are also indicated. They are well-separated on the detector (by about 20 pixels), so that cross-contamination is not an issue. We note that when dispersed, the IFU spectra fall onto the same detector area as spectra from open microshutters, which is why the IFU mode and the multi-object spectroscopy mode (MOS mode) of NIRSpec are mutually exclusive, except for some specific applications\footnote{In low-resolution mode (R100), the IFU spectra occupy only part of the NIRSpec detector area. In principle, it is thus possible to simultaneously obtain spectra through MSA shutters illuminating the unused detector area. However, this combination of IFS and MOS mode is currently not enabled for science observations.}.

   \begin{figure}[h]
   \centering
   \includegraphics[width=\hsize]{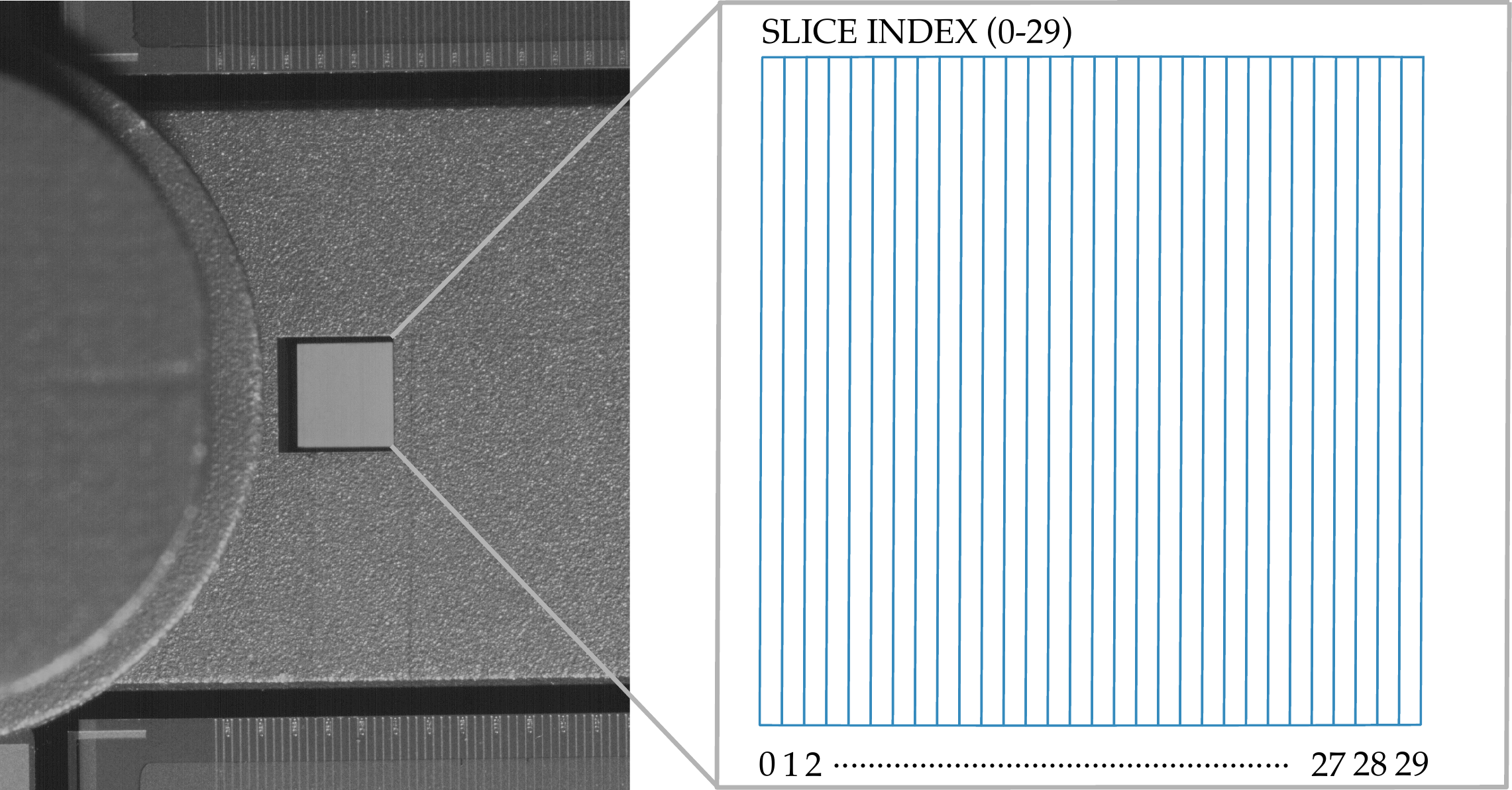}
      \caption{Design and geometry of the NIRSpec IFU aperture. Left: Image of the metal lid (the circular structure on the left) that is mounted on the MSA magnet arm and used to block or unblock the IFU entrance aperture (the square cutout in the center). We note that the picture was taken at ambient temperature, which is why the image slicer does not appear centered in the IFU entrance aperture. Right: Schematic of the image slicer, with slice numbers matching those in Fig. 1.
              }
         \label{fig:lid}
   \end{figure}

The left panel of Fig.\,\ref{fig:lid} shows the square opening through which the pick-off mirror of the IFU optics is illuminated. This aperture can be opened and closed via a metal lid visible on the left. The lid is attached to the magnet arm of the MSA, and by commanding the position of the magnet arm, the IFU optical path can be blocked when NIRSpec is in MOS mode. Conversely, when NIRSpec is in IFS mode and the IFU aperture is open, all microshutters must be commanded closed, as otherwise their light would contaminate the IFU spectra, as discussed in more detail in Sect. \ref{sec:strategies}. Figure\,\ref{fig:lid} also shows the orientation of the image slicer, and the slice numbering scheme that matches the one in Fig.\,\ref{fig:fov}.

\begin{table}[htb]
\caption{Properties of the NIRSpec IFU assembly.}             
\label{tab:ifu_properties}      
\centering                          
\begin{tabular}{l l}        
\hline                
   throughput & $> 50$\% from 0.6 - $3.0\mum$ \\
    & $> 80$\% above $3.0\mum$ \\
    wavefront error & $<$ 100\,nm \\
    stray light & $< 5$\% cross-cont. between slices \\
    operating temperature & 30\,K and above \\
    material & (gold-coated) aluminum \\
    max. dimensions & $140\mm \times 71\mm \times 204\mm$ \\
    total weight & $<$ 1 kg \\
\hline                                   
\end{tabular}
\end{table}

When the IFU aperture is open, light enters the IFU optics, and a $3.1\as \times 3.2\as$ area on the sky is optically rearranged into a line of 30 individual slit images, as illustrated in Fig.\,\ref{fig:fov}. There are two groups of 15 slits, separated by a gap in order to allow for the obscuration by the pick-off mirror structure. This ``virtual long slit'' image is projected back into the remaining NIRSpec optical path such that it will enter the spectrometer optics exactly as if it originated from the MSA image plane, for example from a microshutter or any of the fixed slits shown in Fig.\,\ref{fig:fov}.

\subsection{Optical design}
The design, manufacturing, and performance at subsystem level of the NIRSpec IFU optics has been described in detail in a number of technical papers \citep{clo08,lob08,pur10}. Table\,\ref{tab:ifu_properties} summarizes the key requirements enabled by the design, which is based on the advanced image slicer concept \citep{Con97}, and relies on the free-form capabilities offered by 5-axis diamond machining of aluminum optics.

\begin{figure}[h]
   \centering
   \includegraphics[width=\hsize]{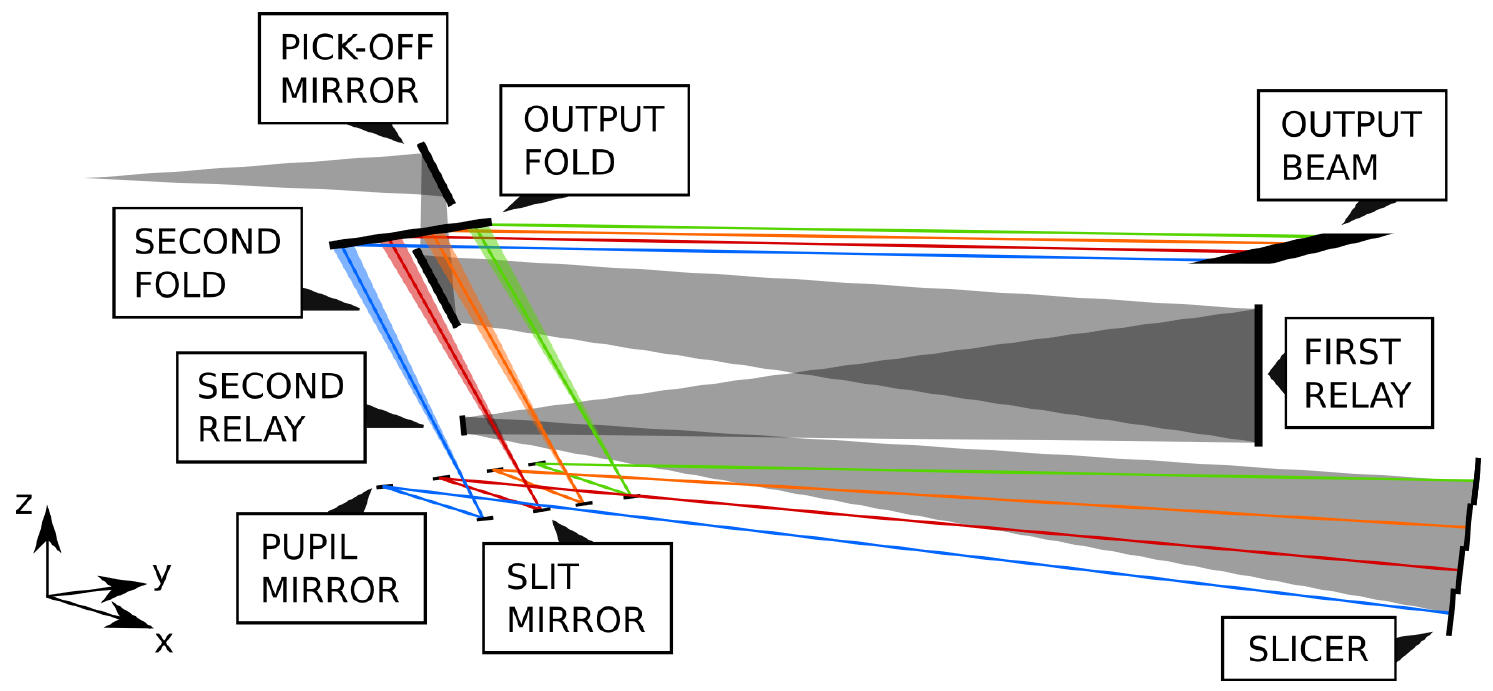}
      \caption{Optical path of the NIRSpec IFU. For clarity, only 4 of the 30 pairs of pupil and slit mirror are shown.
              }
         \label{fig:optical_path}
   \end{figure}

The optical path of the IFU, illustrated in Fig.\,\ref{fig:optical_path}, is based on a folded design concept to guarantee the required compactness (see Table\,\ref{tab:ifu_properties}). Via a flat pick-off mirror and a second fold mirror, the light is directed onto two reimaging (relay) mirrors, which form a magnified image of the input field on the so-called image slicer, which spatially separates the 30 virtual slits. We note that the magnification in the spectral direction is twice as high as that in the spatial direction. This anamorphic magnification is required to guarantee Nyquist sampling of the spectral resolution element. Further details are available in \cite{pur10}.

\begin{figure}[h]
   \centering
   \includegraphics[width=\hsize]{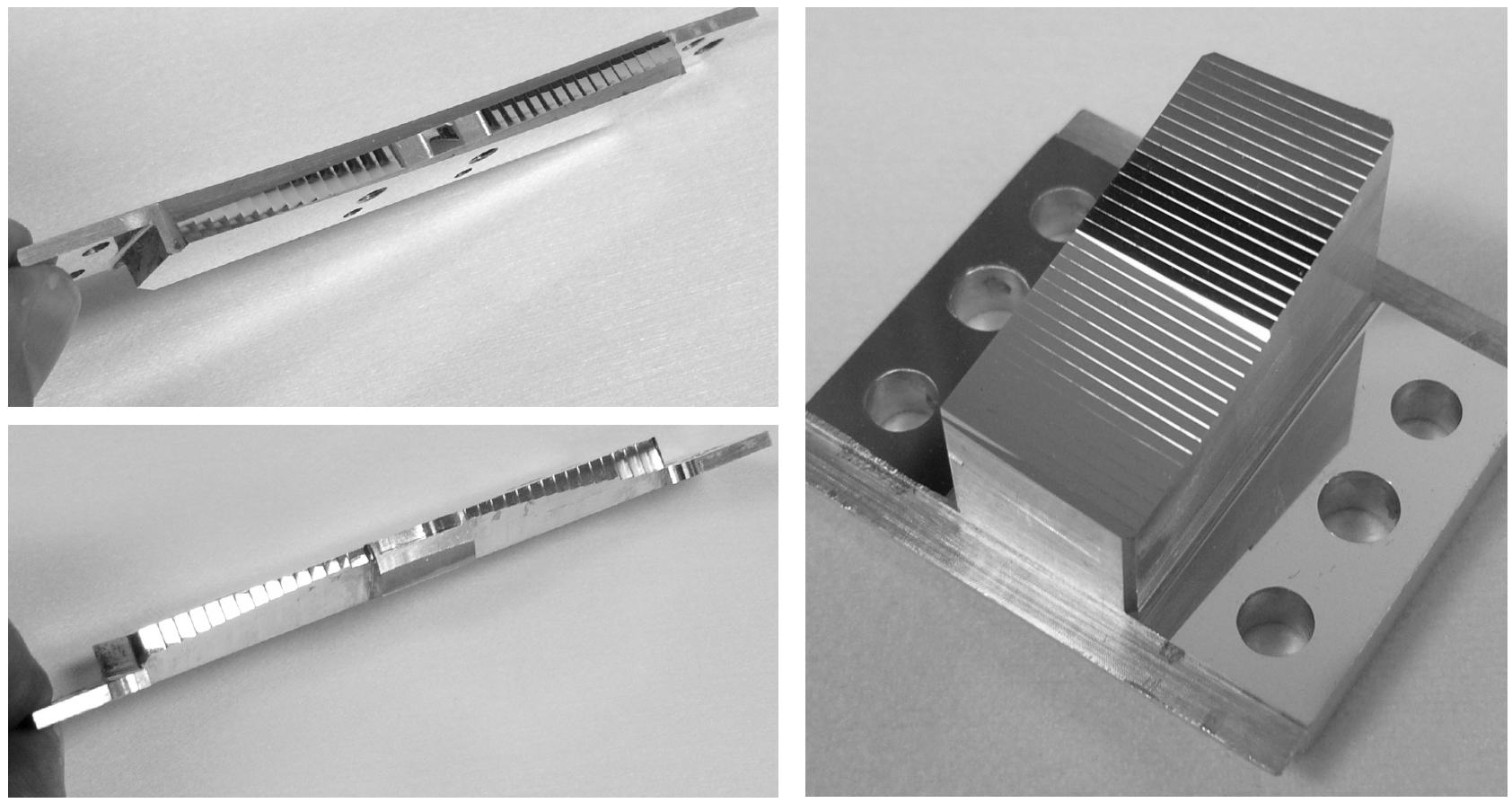}
      \caption{Mirror arrays of the NIRSpec IFU (clockwise from right: slicer, slit mirrors, and pupil mirrors, with relay mirror 2 at the center).  The slicer has an active area of approximately 25 x 12.5mm, while the slit and pupil mirror arrays have an active length of approximately 75mm.
              }
         \label{fig:all_mirrors}
   \end{figure}

The slicer dissects the magnified input field via 30 individual mirror facets, which direct the virtual slit images toward the pupil mirror array. Each slicer facet is individually curved and tilted on two axes to direct the beam toward its dedicated pupil mirror. The 30 pupil mirrors form a single (curved) line, arranged in two groups of 15. We note that the second relay mirror is located between the two groups of pupil mirrors, and is machined on the same substrate, a significant advantage enabled by the free-form diamond machining used to create the IFU optics \citep{lob08}. Both the slicer stack and the assembly of pupil mirrors are shown in Fig.\,\ref{fig:all_mirrors}.

Each pupil mirror is curved and tilted such that the 30 virtual slits are imaged onto a line of slit mirrors. The slit mirrors, also shown in Fig.\,\ref{fig:all_mirrors}, are shaped such that all output beams are directed toward a common plane, the spectrometer entrance pupil. We note that the NIRSpec IFU design differs from other systems in that the slit mirrors cannot be located at the focal plane because of space constraints. They are thus followed by an output fold mirror that reflects the output beam (the virtual long slit of 30 individual slit images) toward the spectrometer optics, where the light is dispersed. 

\begin{figure}[h]
   \centering
   \includegraphics[width=\hsize]{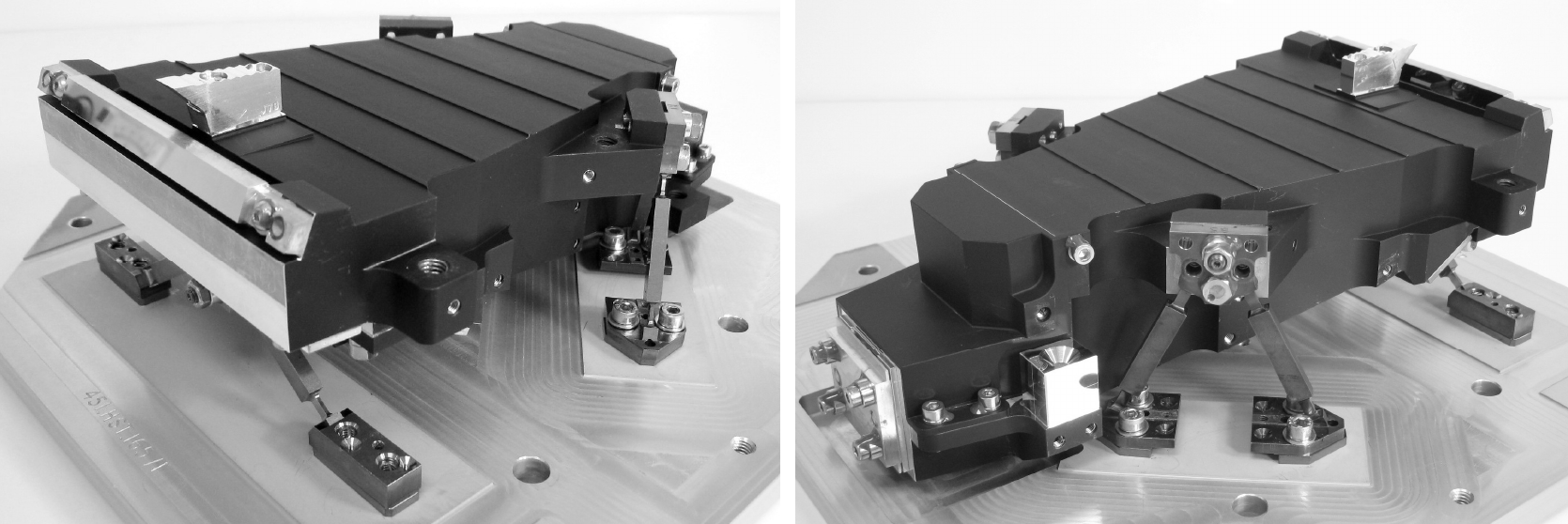}
      \caption{Fully assembled flight model of the NIRSpec IFU (left: front view, right: rear view). The size of the assembly is 140\,mm $\times$ 71\,mm $\times$ 204\,mm.
              }
         \label{fig:assembly}
   \end{figure}

The entire optical path of the IFU is enclosed in a compact, self-contained assembly (shown in Fig.\,\ref{fig:assembly}) that is mounted on the same structure as the NIRSpec MSA.

\subsection{Component-level tests}
The alignment, throughput, and wavefront error (WFE) of the as-built IFU assembly have been tested under cryogenic vacuum conditions, as described by \cite{pur10}. While optical testing of the IFU was limited to wavelengths below $1\,\mum$, the throughput measurements from these tests were fully in line with expectations, with slice throughputs ranging from 0.56 to 0.65 at a wavelength of 700\,nm, and from 0.79 to 0.85 at 905\,nm.

The WFE of all 30 IFU slices was also measured, both at ambient and cryogenic conditions. The results described in \cite{pur10} indicate that the average WFE across the 30 slices is well below the requirement of 100\,nm. While a small number of slices appear to have a slightly larger WFE, this does not degrade the overall NIRSpec WFE requirement (238\,nm), which enables diffraction-limited observations above $3.17\,\mum$.

Lastly, the IFU stray-light performance (i.e., the cross-contamination between slices) was also characterized by illuminating a given slice with a bright point source, and measuring the fraction of light that appears in the two neighboring output slits on either side. The cross-talk measured in this way was between 2.2\% and 3.9\%, which is well within the requirement \citep{pur10}.

\section{Performance of the NIRSpec IFS mode}\label{sec:perf}
\begin{figure*}[t]
   \centering
   \includegraphics[width=0.75\hsize]{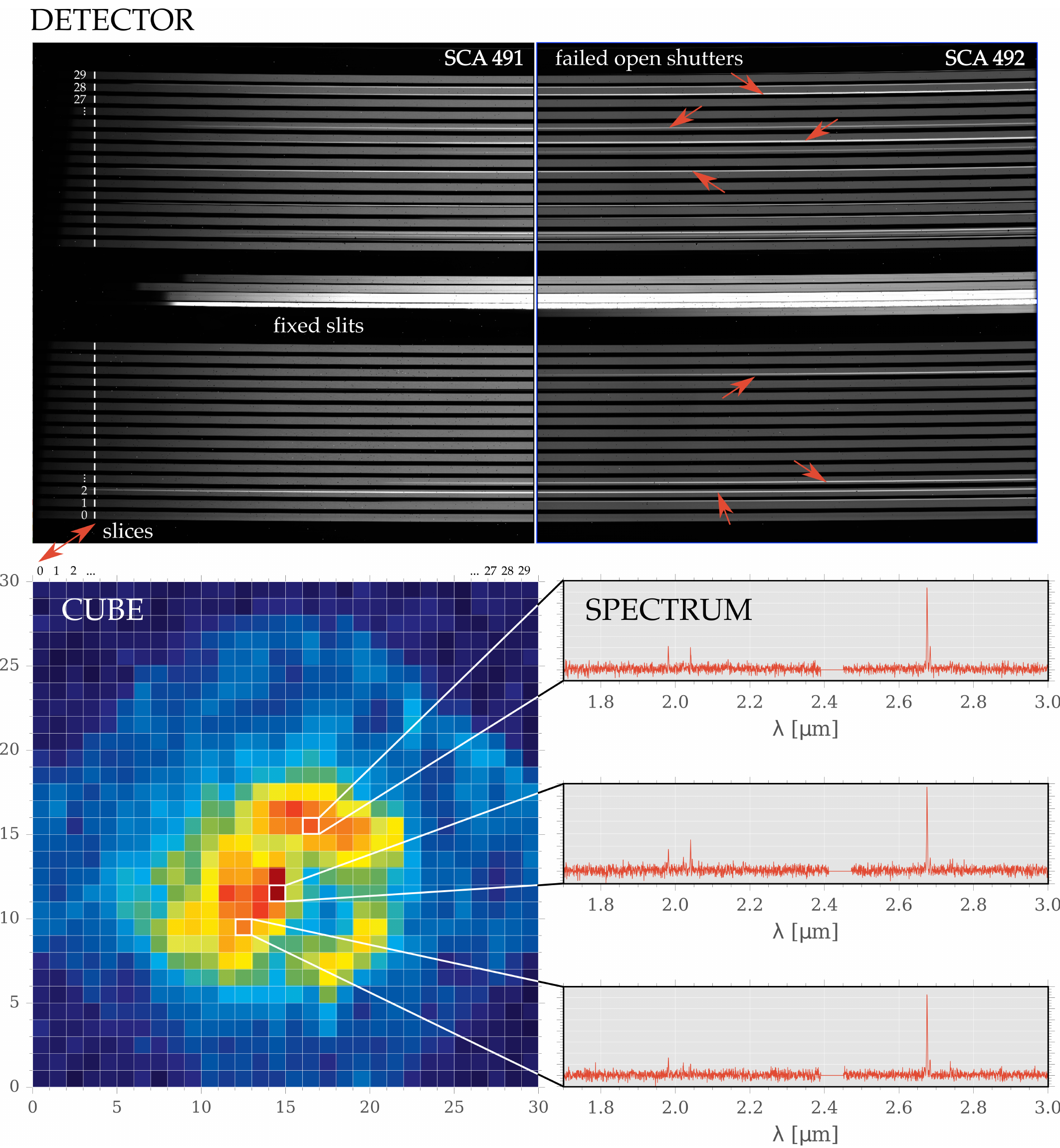}
      \caption{Data products of the NIRSpec IFS mode. Top: Example count rate map (obtained with internal lamp illumination during ground testing), showing the two NIRSpec detectors side-by-side. The numbering scheme of the 30 slices is indicated, as is the location of the ever-present fixed slit spectra. Also highlighted are a number of ``contamination'' spectra caused by failed-open MSA shutters. Bottom left: Example ``channel map'' of a reconstructed data cube (created from simulated science data). The individual slices are identified by number as in the top panel. Bottom right: G235H spectra (extracted from the simulated data cube) for three arbitrary spaxels, illustrating the gap in wavelength coverage caused by the physical separation between the two detector arrays. We note that the exact gap position varies between slices, as described in Sect. \ref{sec:strategies}.
              }
         \label{fig:cube}
   \end{figure*}

The advanced optical design of the NIRSpec IFU described in the last section enables the high-quality reconstruction of the target morphology. As illustrated in Fig.~\ref{fig:cube}, the dispersed images of the 30 IFU slices can be extracted individually, and recombined into a data cube that contains a channel map for each spectral resolution element. The following subsections discuss some performance aspects to be expected for NIRSpec IFU data cubes obtained in orbit, based on ground-test data as well as optical modeling of the instrument and telescope optics. Unless otherwise stated, all numbers are for the entire optical path of the JWST/NIRSpec hardware, that is, including all reflections, filters, gratings, and detectors, as well as any diffraction losses along the way.

   \begin{figure}[h]
   \centering
   \includegraphics[width=\hsize]{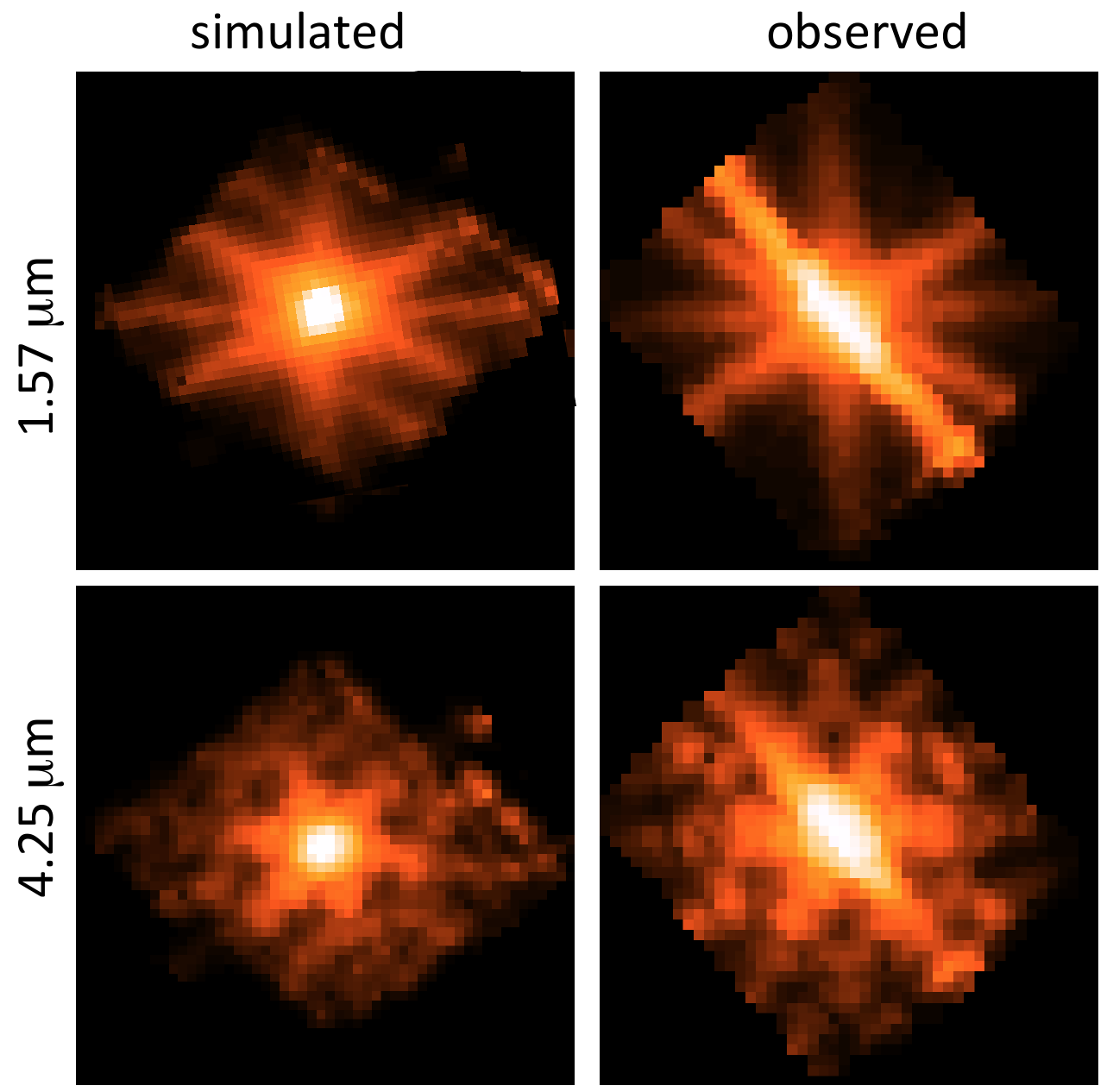}
      \caption{Comparison of simulated (left) to observed (right) point source images in NIRSpec IFU data cubes, showing good agreement in the fainter structures of the diffraction pattern, both at short (top) and long (bottom) wavelengths. Quantitative evaluation of the core width is hampered by the bright diagonal ``bar'' in the observed data, which is an artifact caused by detector saturation.}
         \label{fig:psf}
   \end{figure}

\subsection{Image quality}
In principle, the optical quality of the as-built IFU can be evaluated from a comparison of actual point source observations to predictions from optical models. In practice, this comparison is hampered by the limitations of the optical ground-support equipment (OGSE) in producing realistic point sources at the faint flux levels of JWST science data without the actual telescope.

The best NIRSpec test data to attempt such a comparison between optical models and actual IFU data were obtained during the third cryogenic test campaign of the JWST Integrated Science Instrument Module (ISIM CV3), which took place at NASA's Goddard Space Flight Center in the winter of 2016. The Optical Simulator (OSIM) was used to steer a bright point source onto the IFU entrance aperture, and an exposure was obtained through the CLEAR filter and the PRISM disperser, producing a low-resolution data cube over the wavelength range $0.6-5.3\mum$. This observed data cube can then be compared to a simulated one in the same configuration. To produce the simulated data cube, we used the NIRSpec Instrument Performance Simulator (IPS) described in \cite{piq08,piq10}.

The comparison between the simulated and observed cubes is illustrated in Fig.~\ref{fig:psf} for two channel maps (at $1.57\mum$ and $4.24\mum$). Unfortunately, the OSIM source was too bright to avoid saturation of the NIRSpec detector, which causes ``charge bleeding'' into neighboring pixels. This effect is obvious in the images on the right which show a bright bar-like feature which basically is an image of the IFU slice that contains the PSF core. This prevents a direct comparison of the core width of the PSF. Nevertheless, the good match in the shape of fainter structures in the diffraction pattern indicates that the optical quality of the as-built IFU matches well the theoretical expectations.

As explained in Paper I, the NIRSpec design was optimized for sensitivity, rather than for spatial resolution. As a consequence, the angular size of the NIRSpec detector pixels under-samples the PSF over most of the wavelength range. For full recovery of the JWST spatial resolution in NIRSpec IFS observations, dither strategies need to be employed, which are further discussed in Sect. \ref{subsec:dithers}.

\subsection{Sensitivity}
\begin{figure}[h]
   \centering
    \includegraphics[width=\hsize, trim = {0 0.3cm 0 0.8cm}, clip]{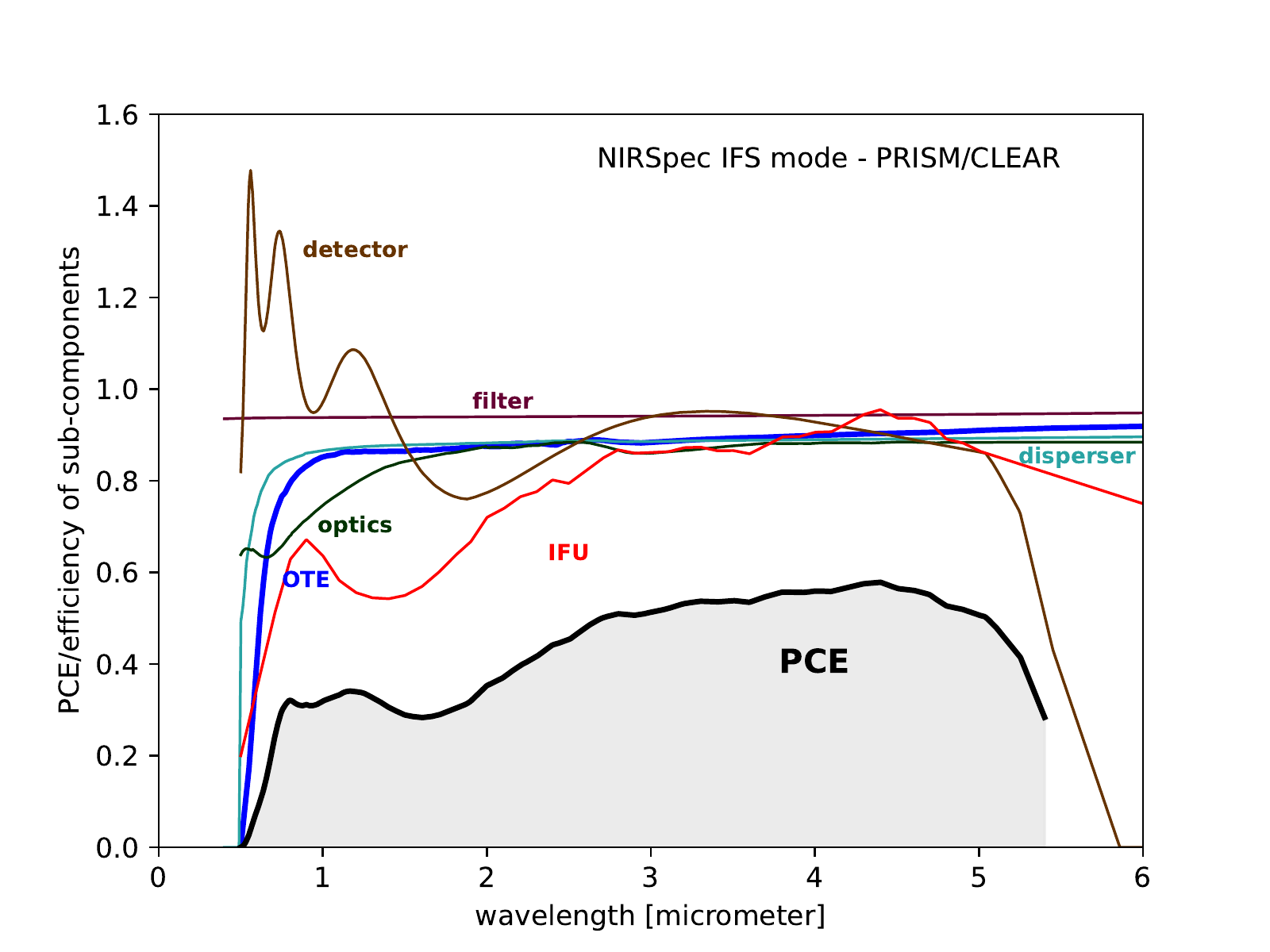} 
    \caption{Photon conversion efficiency as a function of wavelength for the NIRSpec IFS mode with the PRISM disperser and the CLEAR filter (gray-shaded curve). The six main contributors to the total throughput (detector, filter, disperser, and the optical paths of Optical Telescope Element (OTE), NIRSpec, and the IFU itself) are plotted separately. }
         \label{fig:pce_prism}
   \end{figure}

\begin{figure*}[h]
   \centering
  \includegraphics[width=0.48\hsize, trim = {0 0.3cm 0 0.8cm}, clip]{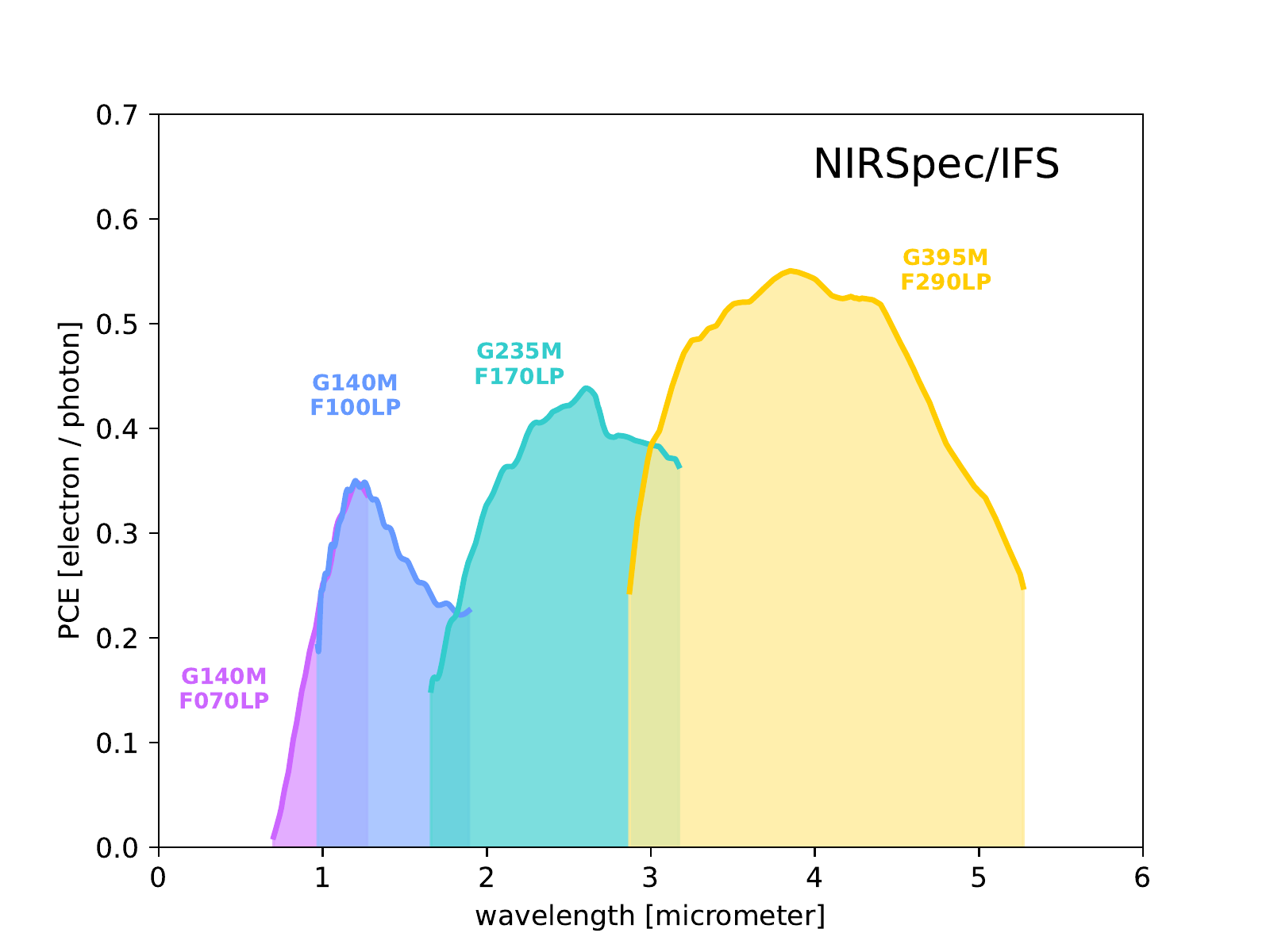}
  \includegraphics[width=0.48\hsize, trim = {0 0.3cm 0 0.8cm}, clip]{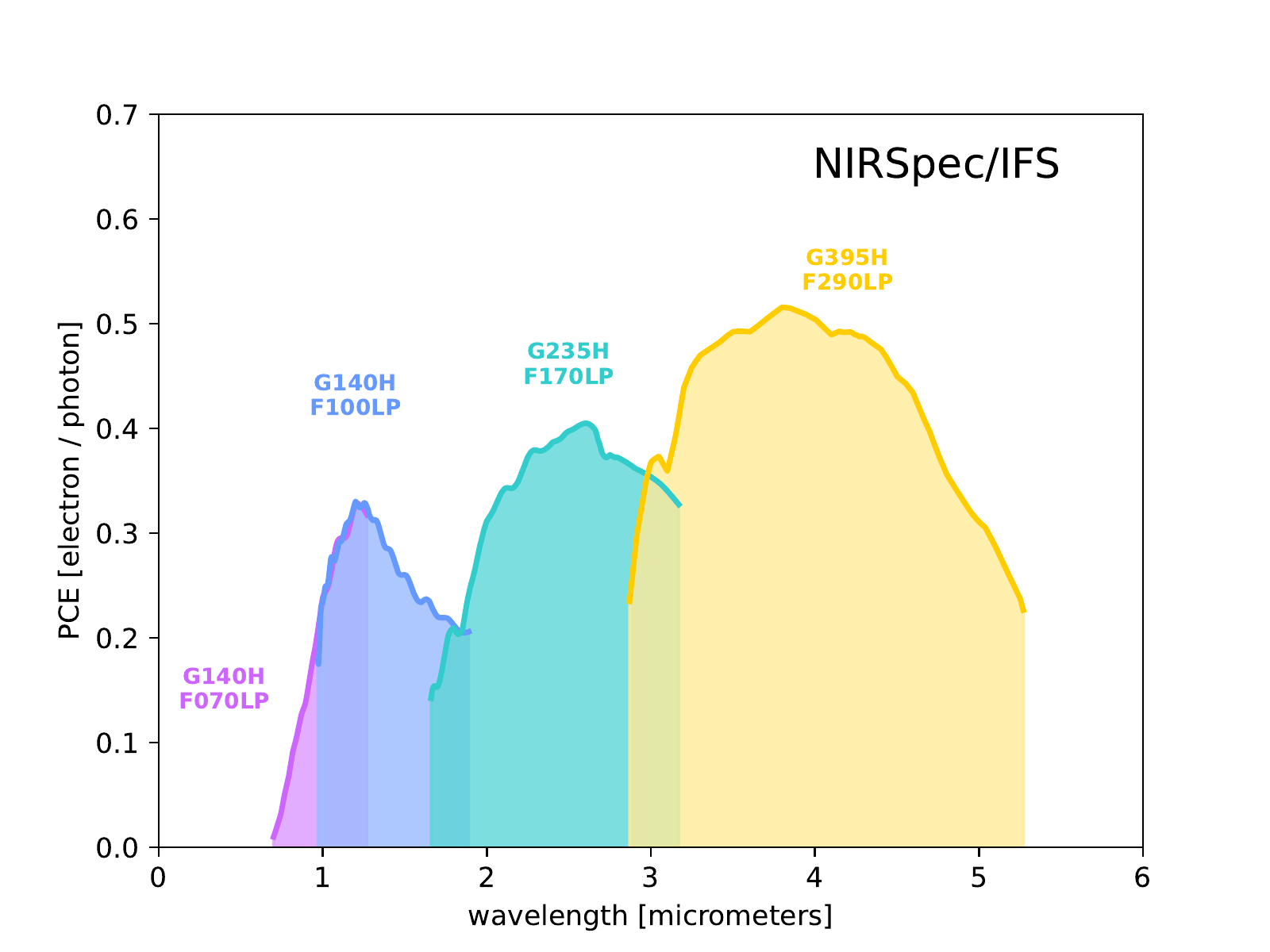}    
\caption{Photon conversion efficiency as a function of wavelength for the NIRSpec IFS mode with the medium- (left) and high-resolution gratings (right). These curves were calculated using the same contributions as shown in Fig. 8.}
         \label{fig:pce_gratings}
   \end{figure*}

\begin{figure*}[h]
   \centering
   \includegraphics[width=0.48\hsize, trim = {0 0.3cm 0 0.8cm}, clip]{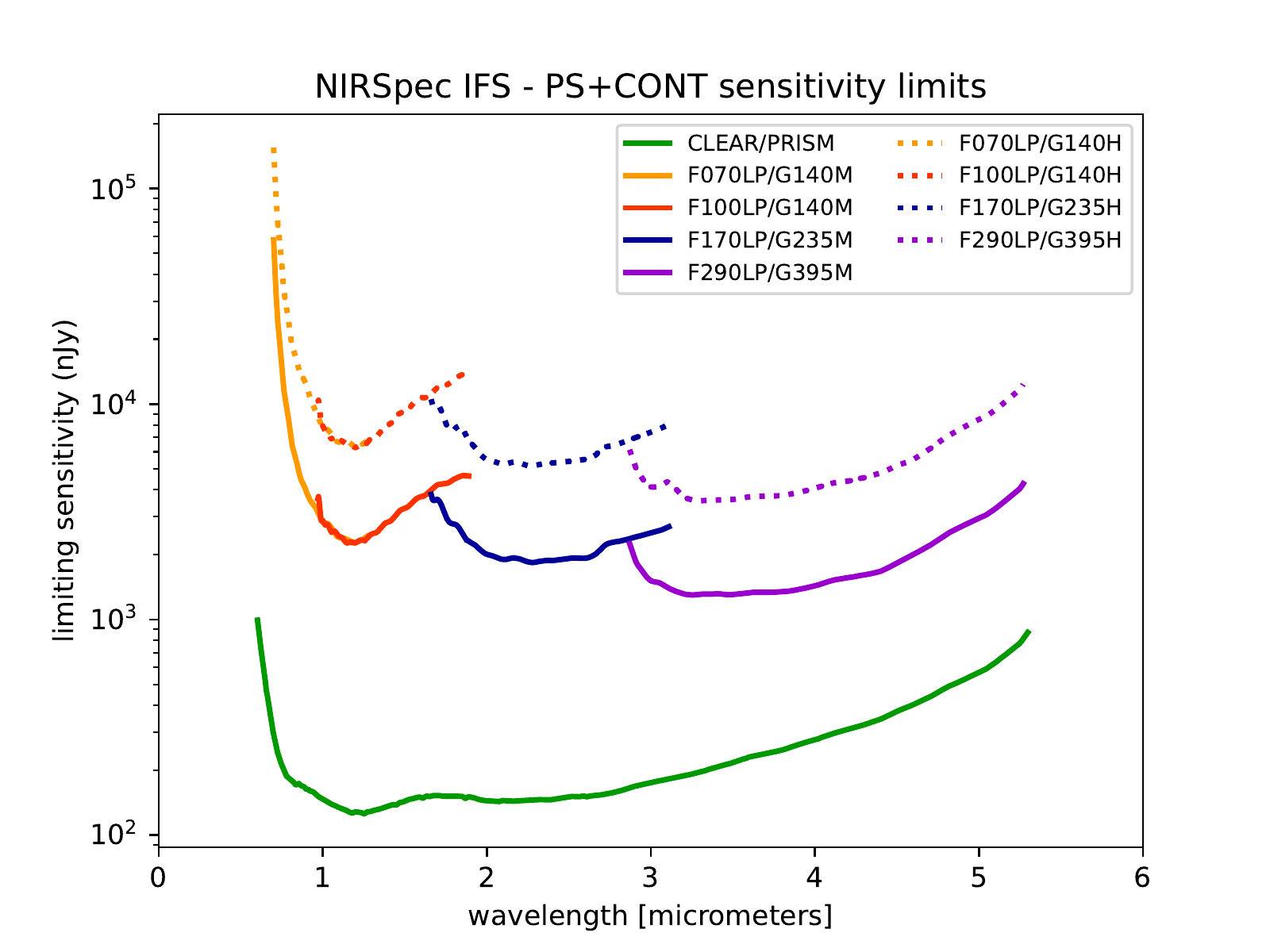} 
   \includegraphics[width=0.48\hsize, trim = {0 0.3cm 0 0.8cm}, clip]{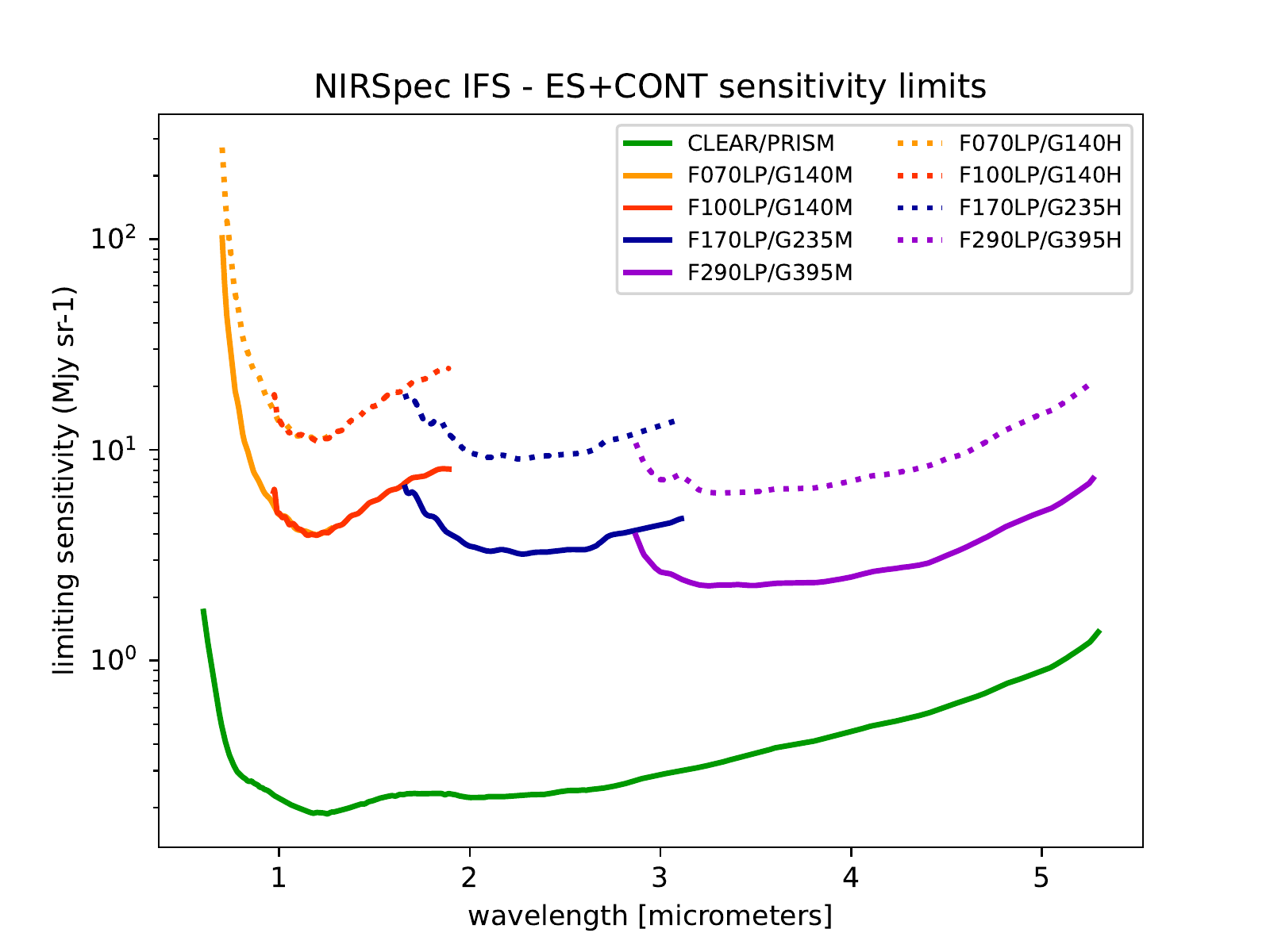} 
   \caption{NIRSpec continuum sensitivity in IFU mode for a point source (left) and uniform extended source (right) observed with the various dispersers. All are for a signal-to-noise ratio of 10 per pixel in 10,000s of on-source integration time.}
         \label{fig:sensitivity}
   \end{figure*}

As discussed in detail in Paper I, the throughput of the optical path is the driving factor determining the photometric sensitivity of the various science modes of NIRSpec. Given that the IFS mode includes an additional eight gold-coated reflections (see Fig.~\ref{fig:optical_path}), one can expect its photon conversion efficiency (PCE) to be somewhat lower than for the other NIRSpec modes. Based on predictions from component-level measurements, and calculated with the same methodology and assumptions as outlined in Paper I, the throughput of the IFU optical path alone is indicated by the red line in Fig.~\ref{fig:pce_prism} which shows the PCE of the IFS mode when using the PRISM/CLEAR configuration. When combined with the other contributing components such as the detector\footnote{As explained in \cite{Rau14}, the detector PCE can be  $> 1$ at wavelengths below $2\mum$ due to the fact that a single photon can excite multiple electron-hole pairs in the semiconductor.}, telescope, and the rest of the NIRSpec optics, the resulting total PCE is in excess of $\approx$50\% longward of $2.5\mum$, and above $\approx$30\% between of $0.7\mum$ and $2.5\mum$. Figure~\ref{fig:pce_gratings} shows the corresponding results for the NIRSpec gratings. Using these PCE curves, and accounting for diffraction losses as outlined in Paper I, the expected NIRSpec sensitivity in IFU mode for the low, medium, and high spectral resolution configurations is shown in Fig.~\ref{fig:sensitivity}, both for a point source (left panel) and an extended source that uniformly fills the aperture (right panel). The exposure parameters used here are the same as in Paper I: ten on-source exposures of 70 groups in NRSIRS2RAPID mode, for a total integration time of $\approx$10\,ks. 

\section{Observing strategies}\label{sec:strategies}

The wide range of science topics that can be addressed with the NIRSpec IFS mode, combined with the many unique aspects of space-borne IR IFS in general, and the NIRSpec instrument design specifically, make a comprehensive discussion of observing strategies challenging. The detailed selection of observation parameters will depend on the properties of the science target, as well as the specific science questions to be addressed. To illustrate this point, we highlight a few specific science cases in Sect. \ref{sec:science}. Here, we only discuss some high-level considerations that JWST users should keep in mind when designing IFS observations with NIRSpec. 

\subsection{Target acquisition}\label{subsec:ta}
Given the relatively small field of view of the NIRSpec IFS mode ($3.1\as \times 3.2\as$), it is prudent to briefly discuss the available methods to position a science object within the IFU aperture. In many (perhaps most) cases, the expected pointing accuracy of JWST without a dedicated target acquisition procedure ($0.15\as$ $1\sigma$ radial) will be sufficient to acquire compact objects, and to keep them within the aperture even when executing the most common dither patterns. 

If the science target is moderately extended ($\approx 1\as -2\as$) and if dithering within the IFU is required, it may be desirable to start the dither pattern at a more precisely defined position in order to keep the target fully within the IFU aperture at all positions. In this case, the first choice should be to use the wide aperture target acquisition (WATA) method, which places the target (or, in the case of faint science targets, a nearby bright reference target with a known offset) in the S1600A slit aperture used for bright object time series observations (see Paper IV), measures its position, and then computes and executes the small-angle maneuver necessary to position it precisely in the IFU aperture.

Lastly, if the target morphology is too complex (i.e., if it lacks a clearly defined emission peak) for the WATA method to work, and if no suitable reference target is available nearby, it is also possible to use the target acquisition procedure for the MOS mode discussed in Paper II. In this case (which comes with significant overheads), the user needs to identify a number of reference stars, and follow the procedure outlined in Sect.\,4.2 of Paper II.
\subsection{Subtraction of sky background}\label{subsec:background}
In addition to the signal from the science target itself, NIRSpec IFS exposures contain flux from a number of unwanted sources that can adversely affect the quality of science data. Examples for such unwanted background emission within the IFU aperture itself are, for example, zodiacal light and/or thermal emission from warmer parts of the observatory. Many JWST science targets will be of comparable brightness to or even fainter than the zodiacal emission alone, and accurate subtraction of the accumulated background signal therefore is essential to reliably derive the target spectrum. 

In this ``faint science'' case, one should apply a ``nodding'' strategy to remove the background signal by differencing subsequent exposures with the science target at different locations. This approach assumes that (i) the background emission is flat across spatial scales comparable to a few times the IFS field of view, and (ii) the background emission is stable at least over the timescale of the observation. Both of these assumptions are likely true, but need to be verified in orbit.

If the source is significantly smaller than $3\as$, the nodding amplitude can be small enough that the target remains within the IFU aperture, so that the full exposure time is spent ``on source.'' For more extended sources, using a larger nodding amplitude is unavoidable, which will move the science target out of the IFU aperture and thus results in less efficient observations. The Astronomer's Proposal Tool\footnote{https://www.stsci.edu/scientific-community/software/astronomers-proposal-tool-apt} (APT) used to prepare JWST observations offers predefined templates for either case. For a more detailed discussion, we refer to the JWST online documentation system JDox\footnote{https://jwst-docs.stsci.edu/}.

Alternatively, a ``master background'' spectrum can be generated and subtracted from all spatial elements in an IFU data cube (hereafter referred to as ``spaxels''). For example, spectra from the fixed slits (which are sky-illuminated in all IFS exposures) or from IFU spaxels known to only contain the background signal can be used to generate a master background
from the on-source science exposure itself. This option, the accuracy of which will be assessed from early on-orbit IFS data, would avoid the efficiency losses inherent to taking dedicated off-source background exposures. 

\subsection{Dither strategies}\label{subsec:dithers}

The NIRSpec IFS mode offers a number of dedicated pointing sequences for the telescope, which vary the target location within the IFU aperture and on the detector. The use of such ``dither'' patterns has a number of benefits: (i) it allows the correction of detector effects such as bad pixels; (ii) it improves the spatial and spectral resolution of the data by providing a better pixel sampling of either resolution element; and (iii) it improves the spectrophotometric accuracy by averaging over errors in the flat-field calibration, as well as path loss differences between slices. Dithering may also help to recover spectral features lost in the gap between the two detectors (see \S\ref{subsec:gap}). 

\begin{figure*}[h]
   \centering
   \includegraphics[width=\hsize]{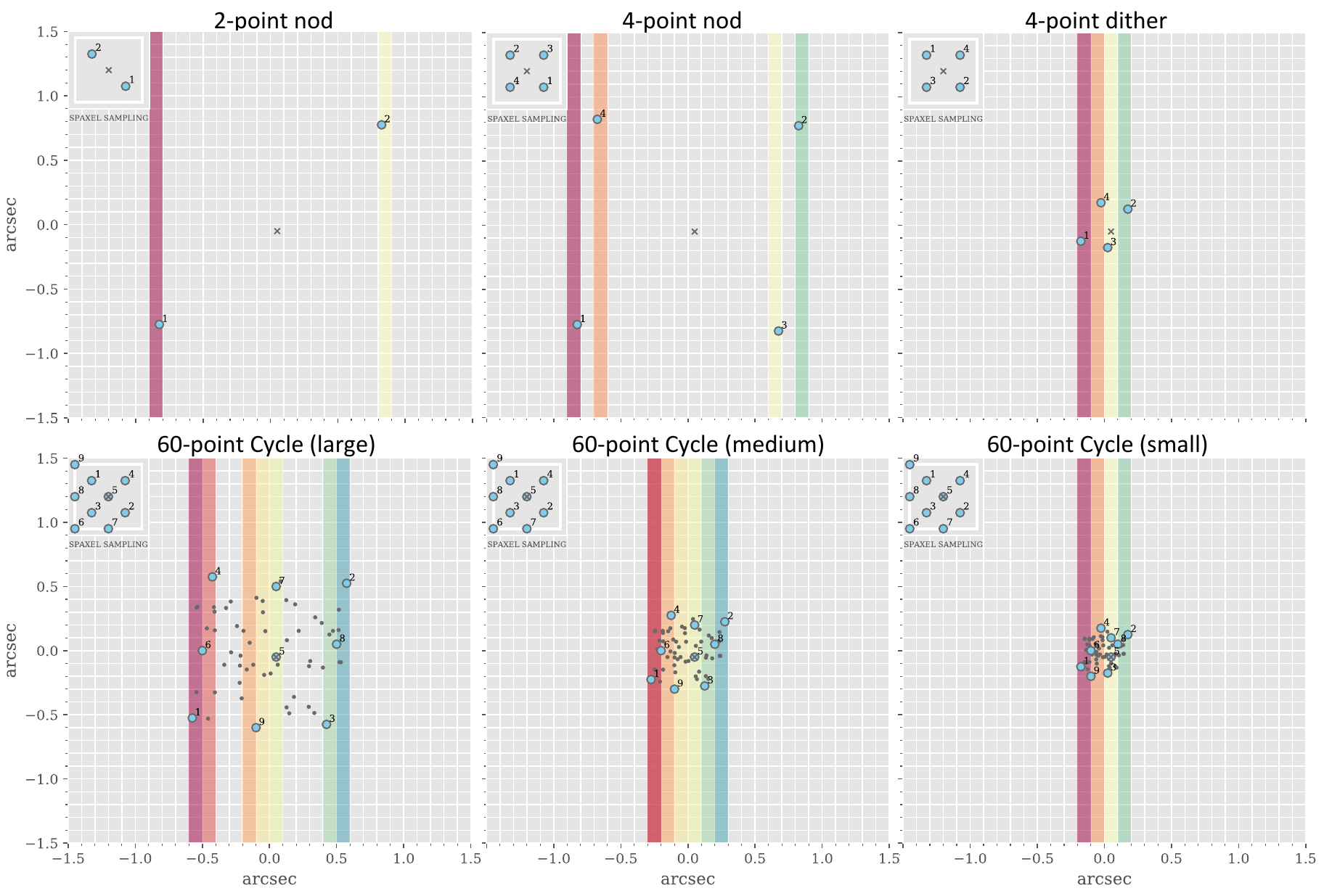}
      \caption{Available nod and dither patterns for the NIRSpec IFS mode. Colored vertical bars mark the slices used for the primary dither points, which are indicated in the top-left corner of each panel, illustrating the spatial sampling of a spaxel. The two-point and four-point nod patterns are mostly used to move compact sources within the aperture for in-field background subtraction. The four-point dither pattern maintains the same spaxel sampling as the four-point nod but is optimized for extended sources that can only tolerate small movements within the aperture. 
      The 60-point cycling patterns allow the user to select an arbitrary subset out of 60 predefined points, with the first nine positions (large symbols) optimally sampling the spaxel. The optimal choice of pattern scale (large, medium, or small) depends on the size of the source.}
         \label{fig:dithers}
   \end{figure*}

Figure \ref{fig:dithers} shows the pointing sequences available in APT for the NIRSpec IFS mode. All available patterns include a sub-spaxel dither to better sample the PSF and therefore improve the spatial resolution of the reconstructed data cube. As mentioned above, the JWST PSF is under-sampled in NIRSpec IFS mode, but the full spatial resolution can be restored by combining dithered exposures, for example by applying the drizzle technique \citep{FH02}. 

The optimum choice of dither pattern depends on target size and background intensity. If the science target is compact (i.e., less than $\approx 1\as$ in diameter), the two-point and four-point nodding patterns are optimal, as they enable background subtraction and improved spatial sampling without loss of on-source exposure time. For more extended sources that (nearly) fill the IFU, the four-point dither pattern restores full PSF sampling while keeping the source well-centered inside the IFU aperture. As discussed in the previous section, in this case, a dedicated ``off-source'' exposure may be necessary for background removal, unless one is in the ``bright science'' case for which the background signal can be ignored.

If recovering the best spatial resolution is the primary goal, three different ``cycle'' patterns (small, medium, and large) are offered within APT. The choice of spatial scale and number of points to be used depends on the target size and brightness. For example, very bright targets that require many short exposures to avoid detector saturation will benefit from using one of the 60-point Cycling patterns.

\subsection{MSA leakage removal} \label{subsec:leakage}
In addition to the in-field background signal discussed in Sect. \ref{subsec:background}, another source of parasitic signal in NIRSpec IFS exposures is caused by the fact that spectra from the IFU slices occupy the same detector area as spectra from the MSA. Therefore, they will be contaminated by any light ``leaking'' through the MSA shutters, for example through failed-open or only partially closed MSA shutters. Here, we briefly explain the nature of this complication, and possible strategies for dealing with it.

\begin{figure*}[t]
   \centering
   \includegraphics[width=0.95\hsize]{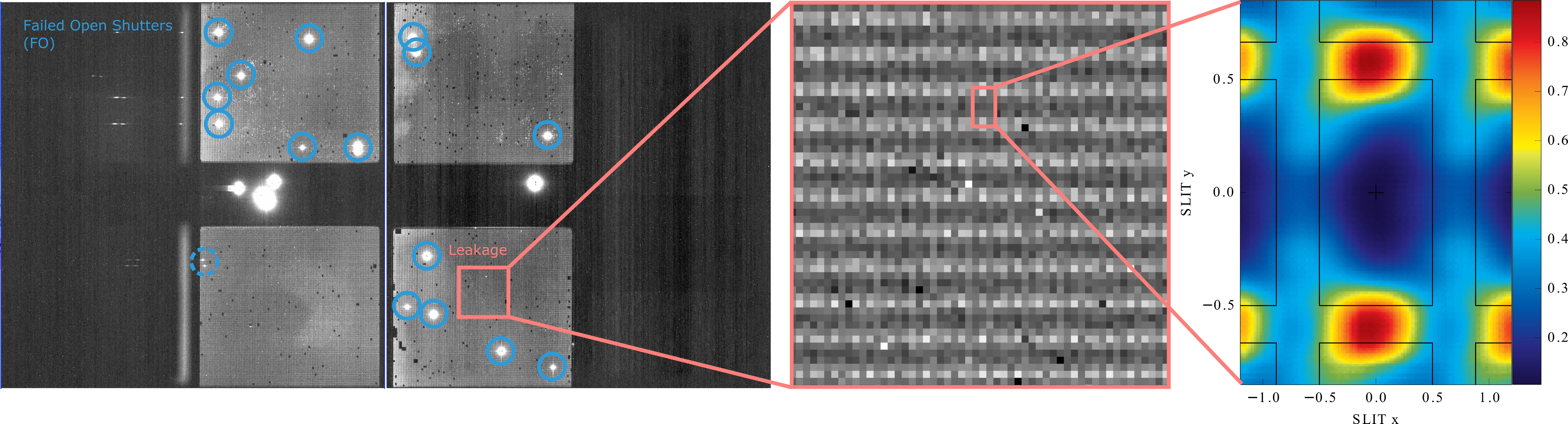}
      \caption{Illustration of the low-level contamination leaking through the closed NIRSpec MSA, as measured from ground-based calibration data. Left: Undispersed exposure through the closed MSA, clearly showing individual failed-open shutters (blue circles), as well as the low-level leakage across the MSA. Center: Enlarged subsection of the MSA showing the leakage pattern around individual shutters. Right: Highly over-sampled image of the leakage pattern, obtained by superposition of the (re-binned) leakage signal from many individual shutters. 
              }
         \label{fig:msa_leakage}
   \end{figure*}

Unfortunately, the NIRSpec MSA is not fully opaque, even when all shutters are commanded closed. Instead, there is low-level leakage through the (closed) shutter grids, as illustrated in Fig.\,\ref{fig:msa_leakage}. While the exact amount of leakage varies across the MSA, typical levels are less than 0.01\% of the signal falling onto the MSA. Nevertheless, this low-level leakage can become significant for many science cases, because when the light is dispersed, the contamination accumulates along the spectral direction, an effect sometimes referred to as ``pileup.'' A quantitative analysis of the impact of leakage pileup on the science performance of NIRSpec IFS observations is presented in \cite{des18}.

Unless this contamination can be ignored (perhaps in the case of a pure emission line source in the IFU), the leakage signal should be subtracted before the creation of an IFU data cube. The cleanest method to achieve this is to obtain a second exposure that is identical to the science exposure in all respects, except that it is taken with the IFU aperture blocked (by moving the lid visible in Fig.\,\ref{fig:lid} across the IFU entrance window). In this way, the ``leakage exposure'' contains only the light reaching the detectors through the MSA (both the pileup and the failed-open spectra). It can simply be subtracted from the science exposure in order to accurately remove the leakage signal, at the cost of somewhat elevated noise levels and additional exposure time. 

For dithered IFU observations, the user has a choice to obtain a leakage exposure either at every dither position or only for the first dither position. The latter approach should suffice in many cases, especially in uncrowded areas of the sky where the leakage signal is not expected to vary on scales of the dither pattern. 

As for the origin of the pileup leakage signal, Fig.\,\ref{fig:msa_leakage} demonstrates that its dominant component has a pinhole-like structure, located on the bars between adjacent shutters. This signal is common to all shutters, and is most likely caused by small gaps in the metal coating that has been put on the  bar structure\footnote{Since the bars are made from silicon, they would be transparent to IR light without such a coating}. Such gaps in the conductive coating are necessary in order to enable individual rows and columns to be addressed and to avoid shorts from affecting large areas on the quadrants. In the future, it may be possible to use this insight to create a leakage model to be subtracted from the science exposure, thus reducing the need for dedicated leakage exposures in orbit.
\subsection{Detector gap and wavelength coverage}\label{subsec:gap}
NIRSpec IFS spectra obtained with the high-resolution gratings (G140H, G235H, and G395H) will have incomplete wavelength coverage, because some wavelengths will fall onto the physical gap between the two NIRSpec sensor chip assemblies (SCAs). In contrast to the multi-object and fixed slit modes (which can move the target to another aperture), the ``lost'' wavelengths cannot be recovered in IFS mode. We note, however, that for compact sources, dithering within the IFU aperture will somewhat reduce the impact of the detector gap. For each of the high-resolution gratings, Table~\ref{tab:gap} lists the affected wavelength range. The exact wavelength range falling into the gap varies from slice to slice, due to the significant slit tilt caused by the out-of-plane illumination of the NIRSpec gratings (see Paper I for details). In addition, there are small differences between exposures, due to the non-repeatability of the grating wheel discussed in \cite{deM12}. The numbers listed in Table~\ref{tab:gap} provide a worst case envelope, in the sense that spectral features that fall within these limits are likely to be missing in at least a portion of the IFU field. We note that the G140H/F070LP configuration is not affected by the detector gap, as its shorter wavelength range is fully captured by a single SCA. The impact of the detector gap on emission-line studies of distant galaxies is further discussed in Sect. \ref{subsec:distant}.

\begin{table}[htb]
\caption{Wavelength coverage of the NIRSpec IFS mode for different spectral configurations. We note that the detector gap only affects observations with the high-resolution gratings (except the G140H/F070LP combination).}             
\label{tab:gap}      
\centering                      
\begin{tabular}{l c c}      
\hline
\noalign{\smallskip}
 Grating/Filter & Science Range & Span of gaps\\
 & [$\mum $] & [$\mum $]\\
\noalign{\smallskip}
\hline                
\noalign{\smallskip}
   PRISM/CLEAR & 0.6 - 5.3 & n/a \\
   G140M/F070LP & 0.90 – 1.26 & n/a \\
   G140M/F100LP & 0.97 – 1.88 & n/a \\
   G235M/F170LP & 1.70 – 3.15 & n/a \\
   G395M/F290LP & 2.88 – 5.20 & n/a \\
   G140H/F070LP & 0.96 – 1.26 & n/a \\
   G140H/F100LP & 0.98 – 1.87 & 1.408 - 1.486 \\
   G235H/F170LP & 1.70 – 3.15 & 2.360 - 2.491 \\
   G395H/F290LP & 2.88 – 5.20 & 3.983 - 4.205 \\
\noalign{\smallskip}
\hline                                
\end{tabular}
\end{table}

\section{IFS data pipeline}\label{sec:pipeline}

   \begin{figure*}[t]
   \centering
   \includegraphics[width=0.95\hsize]{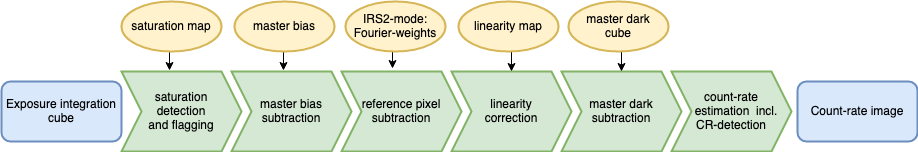} 
   
   \includegraphics[width=0.95\hsize]{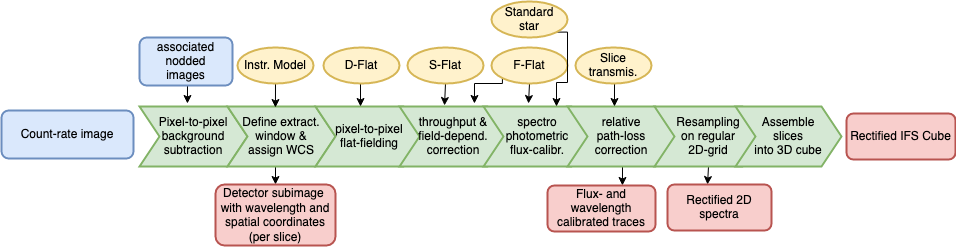}
      \caption{Outline of the data reduction flow for NIRSpec IFS exposures. Top: ``Generic'' preprocessing that converts raw up-the-ramp integrations into count rate maps containing the number of counts per second received by every detector pixel. This portion of the data reduction is common to all NIRSpec observing modes, and in fact all NIR instruments on board JWST. Bottom: Starting with the count rate maps, flow describing the extraction, flux and wavelength calibration, and rectification of each of the 30 slice spectra, so that they can be combined into a three-dimensional data cube per exposure. The combination of multiple dithered exposures into a single ``hyper cube'' is not shown but will be needed to recover full spatial information.
              }
         \label{fig:pipeline}
   \end{figure*}

The overall approach to calibrate the NIRSpec instrument modes has been described in \cite{Boe12}, \cite{Raw16}, and \cite{Alv18}. Here, we briefly describe the data processing steps that are required to turn IFS raw data into three-dimensional data cubes composed of the suite of channel maps, as outlined in Fig.\,\ref{fig:pipeline}. The flow begins with NIRSpec raw exposures, which -- for each of the two NIRSpec detectors -- are essentially cubes of ($2048 \times 2048 \times n$) pixel samples, where $n$ is the number of non-destructive pixels reads ``up the ramp'' of a NIRSpec integration. The preprocessing stage (top panel of Fig.\,\ref{fig:pipeline}) involves the following steps: saturation detection and flagging, master bias subtraction, reference pixel subtraction, linearity correction, dark current subtraction, and count rate estimation. The last step, which is also known as ``slope fitting'', includes ``jump'' detection and cosmic ray rejection \citep[see also][]{Gia19}.

Once the IFS count rate maps have been computed, the spectra of the 30 IFU slices need to be individually extracted, rectified, calibrated, and combined into the final product, that is, the wavelength- and flux-calibrated data cube that contains the spatial information across the full wavelength range of the observation. This nine-step flow is illustrated in the bottom panel of Fig.\,\ref{fig:pipeline}.

The first step is pixel-level background subtraction (when dedicated ``nodded'' background exposures are available and can be subtracted at the level of count rate maps). As discussed in Sect. \ref{subsec:background}, there are other options to correct for the unwanted background emission within the IFU aperture, especially for compact targets. 

The second step is the extraction of the spectral-trace sub-images for each of the 30 slices and the assignment of
  wavelength and spatial coordinates to each pixel therein. This step relies on a high-fidelity instrument software model that allows tracing the light path from the sky through the JWST telescope and NIRSpec optics to the detector array \citep{Dor16}. 

The third is pixel-to-pixel flat-field correction using the detector response reference file (D-flat). The fourth is correction for field-dependent throughput variations, for example between IFU slices, using the reference files for the spectrograph and fore optics response (S-flat and F-flat, respectively).

The fifth step converts the digital detector counts to absolute flux units. The conversion factors as a function of wavelength are derived from observations of spectro-photometric standard stars, with any residual corrections captured in the F-flat. 

The sixth step (correcting for relative path losses) is done differently for point and extended sources. For point sources, this step corrects for path loss differences relative to a centered source. If, on the other hand, the source is marked as extended, the standard pipeline will treat it as a spatially uniform source. In this case, the path loss correction is also calculated relative to a centered point source, and an additional ``pixel-area’' correction is applied to convert the data from flux-per-pixel to surface brightness units. 

The seventh step is a resampling of the 30 detector sub-images onto a common two-dimensional grid of wavelength and spatial coordinates. Finally, the last step is the assembly of a three-dimensional data cube from the 30 resampled sub-images.

The flow depicted in Fig.\,\ref{fig:pipeline} ends with the creation of a rectified data cube for a single IFS exposure. In the case of $n$ dithered exposures (see Sect. \ref{subsec:dithers}), the signal-to-noise ratio and/or the spatial resolution can be improved by combining them into a so-called hyper cube that may have finer spatial sampling than any cube created from a single exposure. This last step is best done using the ($30 \times n$) 2-d spectra directly, in order to avoid multiple instances of resampling of the raw data. In practice, the input detector pixels can be mapped to output cube spaxels using different algorithms, for example a modified Shepard’s method of nearest neighbor-weighted intensity \citep{She68}, or a three-dimensional generalization of the ``drizzle'' technique \citep{FH02} to determine the spatial and spectral overlap between input pixel and output spaxel. In fact, both of these methods will be offered within the JWST user pipeline implemented by the Space Telescope Science Insitute (STScI).

\section{Scientific use cases}\label{sec:science}

We now highlight the scientific potential of the NIRSpec IFS mode, as well as some relevant observing strategies. We do not aim to be comprehensive, and instead focus on a small number of example science cases.

\subsection{Distant galaxies} \label{subsec:distant}

The NIRSpec spectral range encompasses the main rest-frame ultra-violet and optical emission lines for high-z galaxies (with \ha\ observable out to z $\sim$ 7), thus opening up a plethora of diagnostic tools to study the structural, chemical, and dynamical properties of high-redshift galaxies \citep[see, e.g.,][]{Chi20,Nan21}. The fact that each individual NIRSpec grating covers only a fraction of the full NIRSpec spectral range implies that one can optimize the number of spectral configurations to the wavelength range of interest and the redshift of the target. For instance, the \hb\ — \ha\ spectral range is covered with a single grating for the redshift ranges z $\sim$  0.95-1.88 (Band 1), 2.41- 3.83 (Band 2), and 4.90 - 6.92 (Band 3).

The field of view of the NIRSpec IFU aperture ($3.1\as \times 3.2\as$) corresponds to a physical area of $\sim$ 24\,kpc $\times$ 24\,kpc at redshift z=1, and 17\,kpc $\times$ 17\,kpc at z=6. This is generally large enough to cover the entire galaxy emission with sub-kiloparsec spatial sampling, as the spaxel size of $0.1\as$ corresponds to $\sim$ 0.8\,kpc at z=1, and 0.6\,kpc at z=6.  

Because most high-z galaxies have measured angular sizes that comfortably fit within the IFU aperture \citep[e.g.,][]{Bou04,Cur16}, a simple ``point and stare'' approach without a dedicated target acquisition procedure will suffice in many cases, thus avoiding the associated overheads. Also, depending on target size and how smooth the surrounding emission morphology is, an in-field nod or dither approach may be sufficient to enable accurate background subtraction without the need for dedicated off-source exposures.  

Another aspect to consider is the gap between the two NIRSpec detectors discussed in Sect. \ref{subsec:gap}, which affects high-resolution (R=2700) observations and implies that for specific redshift ranges some spectral features cannot be observed over the full field of view. Table~\ref{tab:zgap} shows, for some selected lines, the redshift ranges for which a given line is affected by the detector gap. 

\begin{figure}[h]
   \centering
   \includegraphics[width=\hsize, trim = {0 0.5cm 0 0}, clip]{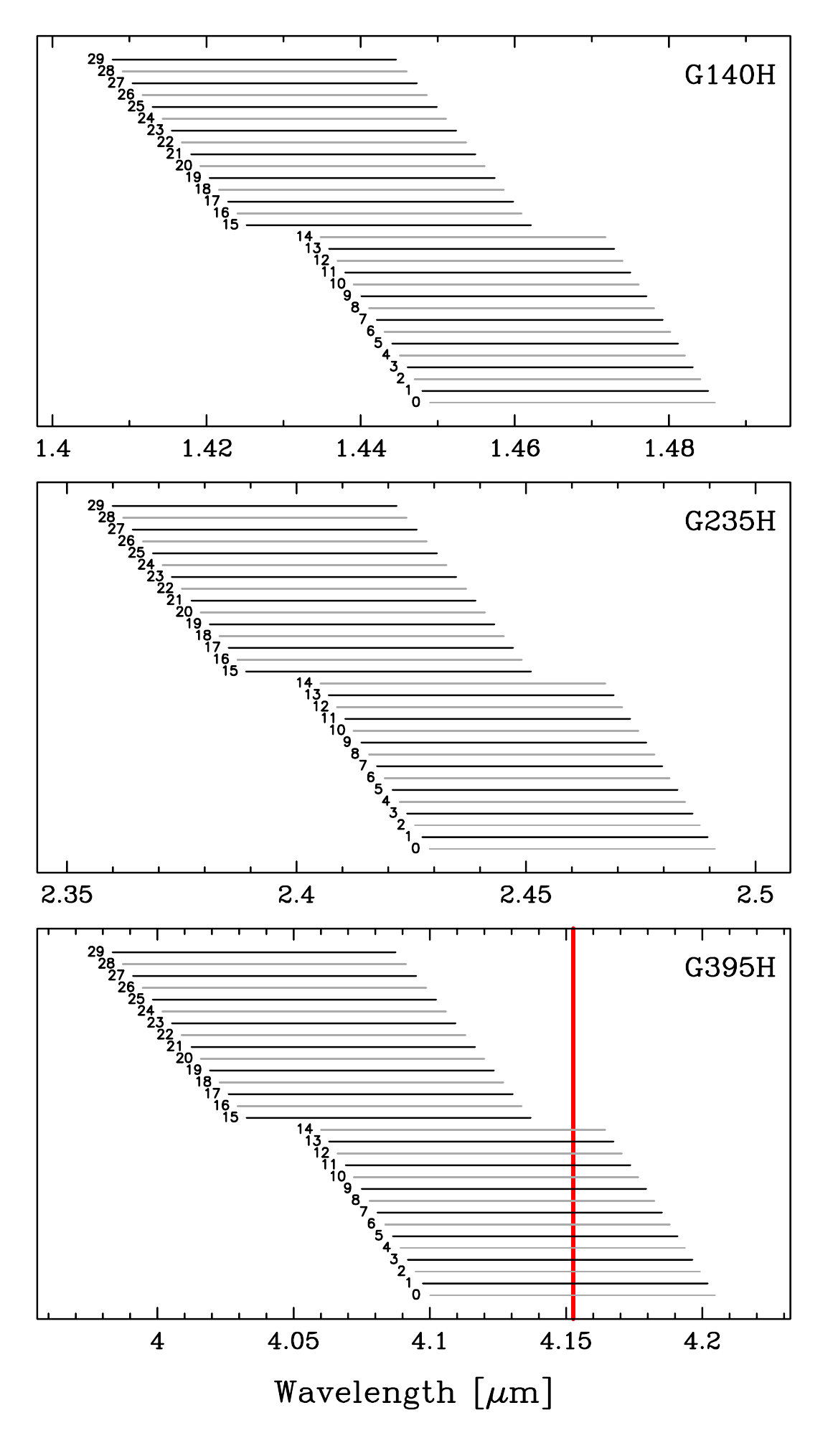}
      \caption{Location of the NIRSpec detector gap for all IFU slices and affected gratings: G140H (top), G235H (middle), and G395H (bottom). For each slice of the IFU, the range of ``missed'' wavelengths is shown by a horizontal bar. The red vertical line in the bottom panel marks the position of the H$\beta$ line for a target at z=7.54. In this case, H$\beta$ cannot be observed across half the field of view. }
         \label{fig:gap_example}
   \end{figure}

It is also important to realise that the precise wavelength range affected by the detector gap depends on the exact location of the target within the IFU aperture. Taking into account that high-z galaxies are in general rather compact, a slight off-centering may therefore allow a particular spectral line to be recovered while still maintaining the whole target emission within the field of view. For example, the bottom panel of Fig.\ref{fig:gap_example}
illustrates the case of G395H observation of a galaxy at z=7.54. The \hb\ line cannot be observed for a group of IFU slices, as it falls on the detector gap for these slices. However, the affected slices only cover about half of the IFU field of view. Therefore, if the object is compact enough, measurements of the H$\beta$ line can be partially recovered by off-centering the target.

\begin{table}[b]
\scriptsize
\caption {Approximate redshift ranges where various emission lines are affected by the detector gap for the different NIRSpec dispersers.}
\label{tab:zgap}
\begin{tabular}{l l c c c}
\hline 
\noalign{\smallskip}
Line  & $\lambda$ [$\mum$]&  G140H & G235H & G395H  \\                           
\noalign{\smallskip}
\hline
\noalign{\smallskip}
Ly$\alpha$&     0.12152          & 10.59 - 11.23 & 18.42 - 19.50 & 31.78 - 33.60 \\
NV        &     0.12408          & 10.35 - 10.98 & 18.02 - 19.08 & 31.10 - 32.89 \\
CIV       &     0.15495          &  8.09 -  8.59 & 14.23 - 15.08 & 24.71 - 26.14 \\
HeII      & 0.16404          &  7.58 -  8.06 & 13.39 - 14.19 & 23.28 - 24.63 \\
CIII]     &     0.19087          &  6.38 -  6.79 & 11.36 - 12.05 & 19.87 - 21.03 \\
MgII      &     0.27991          &  4.03 -  4.31 &  7.43 -  7.90 & 13.23 - 14.02 \\
$[$OII$]$ &     0.37271          &  2.78 -  2.99 &  5.33 -  5.68 &  9.69 - 10.28 \\
$[$OII$]$ &     0.37299          &  2.77 -  2.98 &  5.33 -  5.68 &  9.68 - 10.27 \\
H$\beta$  & 0.48627          &  1.90 -  2.06 &  3.85 -  4.12 &  7.19 -  7.65 \\
$[$OIII$]$&     0.49603          &  1.84 -  2.00 &  3.76 -  4.02 &  7.03 -  7.48 \\
$[$OIII$]$&     0.50082          &  1.81 -  1.97 &  3.71 -  3.97 &  6.95 -  7.40 \\
$[$NII$]$ &     0.65499          &  1.15 -  1.27 &  2.60 -  2.80 &  5.08 -  5.42 \\
H$\alpha$ & 0.65646          &  1.14 -  1.26 &  2.60 -  2.79 &  5.07 -  5.41 \\
$[$NII$]$ &     0.65853          &  1.14 -  1.26 &  2.58 -  2.78 &  5.05 -  5.39 \\
$[$SII$]$ &     0.67183          &  1.10 -  1.21 &  2.51 -  2.71 &  4.93 -  5.26 \\
$[$SII$]$ &     0.67327          &  1.09 -  1.21 &  2.51 -  2.70 &  4.92 -  5.25 \\
CaII      &     0.85004          &  0.66 -  0.75 &  1.78 -  1.93 &  3.69 -  3.95 \\
CaII      &     0.85444          &  0.65 -  0.74 &  1.76 -  1.92 &  3.66 -  3.92 \\
CaII      &     0.86645          &  0.63 -  0.72 &  1.72 -  1.87 &  3.60 -  3.85 \\
$[$SIII$]$&     0.90711$^{\star}$&  0.55 -  0.64 &  1.60 -  1.75 &  3.39 -  3.64 \\
$[$SIII$]$&     0.95332$^{\star}$&  0.48 -  0.56 &  1.48 -  1.61 &  3.18 -  3.41 \\
Pa$\gamma$& 1.09411          &  0.29 -  0.36 &  1.16 -  1.28 &  2.64 -  2.84 \\
Pa$\beta$ & 1.28216          &  0.10 -  0.16 &  0.84 -  0.94 &  2.11 -  2.28 \\
$[$FeII$]$&     1.64445$^{\star}$&  ...          &  0.44 -  0.51 &  1.42 -  1.56 \\
Pa$\alpha$& 1.87561          &  ...          &  0.26 -  0.33 &  1.12 -  1.24 \\
Br$\gamma$& 2.16612          &  ...          &  0.09 -  0.15 &  0.84 -  0.94 \\
\noalign{\smallskip}
\hline
\end{tabular}
\tablefoot{\scriptsize Wavelengths are from 
http://classic.sdss.org/dr6/algorithms/speclinefits.html and NIST (Kramida et al. 2018). 
All are listed for vacuum; those marked with an asterisk were converted from air to 
vacuum using the transformation in \cite{cid96}. }
\end{table}

\subsection{Nearby galaxies}
The NIRSpec IFS mode offers exciting science prospects also for studies of nearby galaxies. The JWST wavelength range, most of which is difficult or even impossible to study from the ground, contains a large number of diagnostic spectral features from stars, dust, and interstellar gas (molecular, neutral, as well as ionized). Given the superior sensitivity and spatial resolution of JWST, IFS studies of nearby galaxies with NIRSpec and MIRI thus promise many breakthrough discoveries, especially in dust-enshrouded environments.

On the other hand, the small size of the IFU aperture may become a limitation, because at a typical distance of 10\,Mpc, it only covers an area of about ($50\times 50$)\,pc$^2$. Most likely, NIRSpec IFS studies of nearby galaxies will therefore be limited to compact regions of interest, for example their nuclear regions. Here, a primary goal is the search for, and characterization of, ``elusive'' active galactic nuclei \citep[see, e.g.,][]{Sat21}. Other compact objects in external galaxies that are well-suited for IFS studies with JWST include extreme starbursts, (embedded) young stellar clusters, or individual sources such as protostars or planetary nebulae. 

Irrespective of the nature of the science target, it is likely that in nearby galaxies, the surrounding emission from the host galaxy's stellar body is the dominant source of background emission. Because this background is likely nonuniform, especially in the often dust-rich nuclei of nearby galaxies, the in-field nodding approach for its subtraction is unlikely to yield good results. Therefore, unless the in-field background emission can be ignored because the science target is much brighter than its surroundings (e.g., in the case of active galactic nuclei), its subtraction should instead be done using either a dedicated background exposure of the empty sky or a `master background' spectrum created from spaxels surrounding the science target. 

An additional complication may come from the fact that the leakage signal of the galaxy body through the MSA shutter grid is potentially non-negligible, and thus needs to be subtracted, possibly even on an exposure-by-exposure basis (i.e., on a per dither basis).

Lastly, some galactic nuclei are bright enough that saturation of the detector pixels becomes an issue even for the shortest integrations. We note that NIRSpec IFS exposures can only be taken in full-frame mode: shortening the frame time by using subarrays is not an option in IFS mode. Quantitative details of the NIRSpec saturation limits can be found in the JWST online documentation system (JDox). 

\subsection{Milky Way and the Local Group}
JWST's IR sensitivity and high spatial resolution offer exciting prospects for understanding the life cycle of dust in the Milky Way and its nearest neighbors. Building on imaging surveys with previous IR space missions \citep[e.g.,][]{Boy15}, integral-field observations with NIRSpec and/or MIRI will yield detailed spectroscopy of individual dust-producing sources, and thus will enable a quantitative comparison of the dust production rates in different types of objects, such as  asymptotic giant branch stars, supernova remnants, planetary nebulae, or red supergiant stars. 

Another important topic certain to benefit from JWST is the spatially resolved study of young stellar objects with their (potentially planet-forming) circumstellar disks, and a number of Cycle 1 programs are planned to observe the protoplanetary and debris disks around Herbig Ae stars, T Tauri stars, or brown dwarfs. Here, integral-field observations with NIRSpec and MIRI will enable a deeper understanding of (i) the chemical inventory in the terrestrial planet forming zone, (ii) the gas evolution into the disk dispersal stage, and (iii) the structure of protoplanetary and debris disks in the near- and mid-IR.

Lastly, JWST IFS observations offer a unique opportunity to penetrate the thick veil of dust toward the supermassive black hole Sgr\,A$^*$ occupying the Galactic center. The main science goals here are to detect and characterize the mid-IR properties of Sgr A$^*$ itself, as well as to study the composition and kinematics of the central stellar cluster at mid-IR wavelengths with unrivaled sensitivity and spatial resolution. The main observational complication for Galactic center studies with NIRSpec is the extremely high background that can cause detector saturation as well as high levels of leakage signal through the MSA (see Sect. \ref{subsec:leakage}), and thus requires careful planning of telescope pointing and exposures parameters.

\subsection{Solar System observations}
The attitude control system of JWST is capable of tracking moving targets with proper motions up to $\rm 30\,mas\,s^{-1}$, and possibly higher. This enables observations of a wide range of targets within the Solar System, as discussed in \cite{Mil16}. In particular, NIR reflectance spectroscopy can reveal the molecular inventory and surface composition of near-Earth objects, asteroids, comets, and trans-Neptunian objects, as well as the outer planets and their moons. Given that most minor bodies in the Solar System have angular sizes that comfortably fit within the NIRSpec IFU aperture, target acquisition requirements are somewhat relaxed, and therefore, this mode is particularly efficient for their study.

Moreover, the NIRSpec wavelength range contains a large number of features from molecules such as $\rm H_2O, CO, CO_2$, and $CH_4$, as well as from various ices and minerals. The NIRSpec spectral resolution, combined with the superior sensitivity of JWST, will allow detailed abundance studies of these, and thus yield a significant advancement in understanding the composition and formation history of the Solar System. A more detailed discussion of the wide range of Solar System science programs enabled by JWST is provided in a special edition of PASP\footnote{PASP, 2016, vol 128, no. 959}.

\section{Summary and conclusions}\label{sec:summary}
The JWST will be the first space astronomy mission to offer IFS at NIR and MIR wavelengths. In this paper we have presented an overview of the design, functionality, performance, and scientific applications of the IFU of the NIRSpec instrument on JWST. We have also outlined some of the considerations that are indispensable in designing a successful and efficient observing program. Given the superior sensitivity and spatial resolution of JWST, NIRSpec IFS studies promise breakthrough discoveries for a variety of science topics, and with a modest investment of exposure time (a few hours for the faintest high-redshift sources, minutes for nearby galactic nuclei).

The final performance of the NIRSpec IFS mode in terms of sensitivity, background subtraction strategies, and pointing accuracy and stability will be evaluated during the JWST commissioning phase, which will be concluded six months after launch.

\begin{acknowledgements}
The NIRSpec IFU was designed and built by Surrey Satellite Technology Ltd (SSTL) in consultation with the Centre for Advanced Instrumentation (CfAI) of Durham University, and under supervision by the NIRSpec prime contractor, Airbus GmbH (formerly EADS Astrium). We are grateful for the efforts of everyone involved. 

SA acknowledges support from the Spanish Ministerio de Ciencia e Innovación through grant PID2019-106280GB-I00.

AJB acknowledges funding from the “FirstGalaxies” Advanced Grant from the European Research Council (ERC) under the European Union’s Horizon 2020 research and innovation programme (Grant agreement No. 789056).
\end{acknowledgements}


\end{document}